\documentclass[a4paper,11pt]{article}
\usepackage{jheppub}

\RequirePackage[T1]{fontenc}

\usepackage{graphicx,xcolor}
\usepackage{mathptmx}      
\usepackage{flushend}
\usepackage{amsmath,amssymb}
\usepackage[normalem]{ulem}
\usepackage{subcaption,arydshln}

\usepackage{lineno}
\usepackage{graphicx} 
\usepackage{amsmath}
\usepackage{amsfonts}
\usepackage{cleveref}
\usepackage{booktabs}
\usepackage{xcolor}
\usepackage{subcaption}
\usepackage[ruled,vlined]{algorithm2e}
\usepackage{color}
\usepackage{colortbl}
\usepackage{tikz}
\usepackage{pgffor}
\usepackage{subcaption}
\usepackage{makecell}

\usepackage{amssymb, physics, multirow, array}

\newcommand{\dR}{\Delta R}
\newcommand{\pt}{p_{\mathrm{T}}}
\newcommand{\softmax}{\operatorname{softmax}}

\definecolor{lime}{HTML}{A6CE39}
\DeclareRobustCommand{\orcidicon}{\hspace{-2.1mm}
\begin{tikzpicture}
\draw[lime, fill=lime] (0,0) circle [radius=0.13] node[white] {{\fontfamily{qag}\selectfont \tiny \,ID}}; \draw[white, fill=white] (-0.0525,0.095) circle [radius=0.007]; 
\end{tikzpicture} \hspace{-3.2mm} }
\foreach \x in {A, ..., Z}{\expandafter\xdef\csname orcid\x\endcsname{\noexpand\href{https://orcid.org/\csname orcidauthor\x\endcsname} {\noexpand\orcidicon}}}

\title{\boldmath This is SParTA: \\
A modular, efficient and scalable transformer architecture for jet-tagging}

\author[a,b]{Rajneil Baruah\orcidA{},}
\author[a]{Subhadeep Mondal\orcidB{},}
\author[b]{Sunando Kumar Patra\orcidC{},}
\author[b]{Satyajit Roy\orcidD{}}

\affiliation[a]{Department of Physics, SEAS, Bennett University, Greater Noida, Uttar Pradesh, 201310, India}
\affiliation[b]{Department of Physics, Bangabasi Evening College, Kolkata, 700009, West Bengal, India}

\emailAdd{rajneilb.physics@gmail.com}
\emailAdd{subhadeep.mondal@bennett.edu.in}
\emailAdd{sunando.patra@gmail.com}
\emailAdd{roy.satya05@gmail.com}

\abstract{We present the Sparse Particle-interaction Transformer Architecture (SParTA), a modular, scalable, and efficient transformer for jet tagging at high-energy colliders. Exploiting the steeply falling transverse-momentum ($p_T$) spectrum of jet constituents, SParTA factorizes constituent processing into two topological regimes. Ordered by descending $p_T$, only the $N$ hardest constituents (the \textit{hard core}) participate in quadratic $O(N^2)$ differential interaction attention driven by pairwise physical and learned relational features. The remaining soft constituents are integrated via linear $O(lN)$ cross-attention to the hard core, preserving soft-radiation and color-flow information while avoiding the $O(l^2)$ memory bottleneck of full self-attention over all $l$ constituents. Pairwise interaction features are computed dynamically in-network from constituent four-momenta and conditioned on global jet-level substructure observables. On the quark-gluon tagging benchmark, a 300K-parameter SParTA model achieves an AUC of 92.02\% and accuracy of 85.06\%—matching the 2.14M-parameter Particle Transformer (ParT) at one-seventh the parameter count and outperforming IAFormer by 21\% in background rejection at 30\% signal efficiency. On the boosted top-tagging benchmark, SParTA outperforms ParT and ParticleNet with 511K parameters, while we analyze the architectural trade-offs relative to fully dense multi-layer interaction networks.}

\begin{document}
\maketitle


\section{Introduction}
\label{sec:intro}

Advent of deep learning \cite{Goodfellow-et-al-2016, Larkoski:2017jix, Guest:2018yhq, Mikuni:2025tar} has revolutionised jet tagging at the large hadron collider (LHC) by moving away from the traditional multivariate methods based on hand-crafted physics features to analyse low order detector data \cite{Thaler:2010tr, Thaler:2011gf, Stewart:2010tn, Larkoski:2013eya, Moult:2016cvt, Komiske:2017aww, Abdesselam:2010pt, Kogler:2018hem, Altheimer:2012mn, Altheimer:2013yza, Adams:2015hiv, Marzani:2019hun, Larkoski:2017jix}. This transformation has drastically improved performance across various jet classification tasks. Early deep learning approaches treated jets as images by mapping energy depositions in calorimeters onto a 2D grid. Models like Convolutional Neural Networks (CNNs) \cite{Baldi:2016fql, Kagan:2020yrm, Kasieczka:2017nvn, Macaluso:2018tck, Kasieczka:2019dbj, Diefenbacher:2019ezd, Bollweg:2019skg, Butter:2017cot, Baruah:2025vpe} achieved significant improvements over traditional methods but faced challenges with data sparsity and difficulty incorporating non-additive information such as particle type of the jet constituents \cite{Qu:2019gqs}. Models later began treating jets as sequences of constituent particles, often sorted by transverse momentum \cite{Guest:2016iqz, Pearkes:2017hku, Egan:2017ojy, Fraser:2018ieu, CMS-DP-2017-013, CMS-DP-2018-058, CMS-DP-2017-049, CMS:2020poo}. However, because jet constituents have no intrinsic order, manually imposed sorting could impair performance \cite{Qu:2019gqs}.

One of the major breakthroughs in the study of jets has been representing jets as a particle cloud. The particle cloud representation treats a jet as an unordered, \textit{permutation-invariant} set of particles (the constituents of a jet) ~\citep{Shlomi:2020gdn, Duarte:2020ngm, Thais:2022iok, Komiske:2018cqr, Qu:2019gqs, Bogatskiy:2020tje, Gong:2022lye, Bogatskiy:2023nnw, Bogatskiy:2022czk, Bogatskiy:2023fug}. This highly flexible method allows models to attribute arbitrary features to every particle and, for some years, has become the foundation for many Graph Neural Network (GNN)-, Lorentz-equivariant-, or transformer-based architectures, which now define the state of the art in jet tagging. 
The transformer-based architectures can be best understood as pursuing this goal along three independent (but not necessarily \textit{orthogonal}) axes, each aimed at reducing computational cost without sacrificing performance: how \textit{symmetry and representation} are incorporated into the network, whether and how pairwise physical interactions are encoded in the attention mechanism, and how the attention mechanism itself is structured or restricted to control its computational scaling.

Particle Transformer ParT was the first in a long line of transformer-based jet tagging architectures that first used pairwise physical features of jet constituents, added directly to the ($QK^{\top}$) term as a bias, before the SoftMax operation. It is termed as Particle-Multi-Head Attention(P-MHA). Using P-MHA blocks over standard self-attention mechanism ParT was able to achieve performance better than ParticleNet~\citep{Qu:2019gqs}.  MIParT~\citep{Wu:2024thh} further builds upon the importance of pairwise features and create a new-mechanism called More-Interaction Attention(MIA), where the embedded pairwise feature tensor itself is used as the attention score instead of using it as a bias, like in P-MHA. They then stack multiple MIA blocks followed by P-MHA blocks and class attention blocks to complete their architecture. IAFormer~\citep{Esmail:2025kii} in its recent work, inspired from the MIA blocks of MIParT use just the pairwise interaction tensor as the attention score. Each layer implements differential attention~\citep{difftransformer} mechanism and transform the sequence and the interactions independently which are used as sequence inputs and attention-score, respectively for the next layer. With number of parameters fewer by orders of magnitude, IAFormer is able to achieve performance which is comparable to that of ParT and ParticleNet.

Among the earliest architectures to use attention mechanism in such works, ABCNet~\citep{Mikuni:2020wpr, Mikuni:2021pou} uses a GAPLayer which constructs local graphs by connecting constituent particle to its k-nearest neighbour in the $(\eta, \phi)$ plane. Then a multi-head attention mechanism repeats the same, but also aggregates graph features as done in a standard GNN and further concatenates with global jet features to enrich the extraction of non-local features. ECT(Edge Convolution Transformer)~\citep{VasquezPlaza:2026vjg} uses edge convolutions over k-nearest neighbours and then performs self-attention over their output representation. This combines the idea of local inductive bias of ParticleNet with the standard attention mechanism of a transformer.The Mixer-Network~\citep{Hammad:2024cae} replaces the quadratically expensive self-attention blocks of standard transformers with and MLP-Mixer network. They re-cluster the jets with standard jet-clustering algorithms~\citep{Cacciari:2008gp, Dokshitzer:1997in, 10.1007/978-3-642-37456-2_14} to get more substructure or \textit{prong} information of the jet and use cross attention between the jet features with the subjet features.  

L-GATr is the Lorentz-Equivariant Geometric Algebra Transformer~\citep{Brehmer:2024yqw} which hardcodes the jet substructure into the network by representing particles as multivectors in spacetime geometric algebra. Since all operations act on these multivector representations, the network remains equivariant under Lorentz transformations. 

Note that, the standard self-attention mechanisms scale quadratically with the number of constituent particles in a jet. This imposes severe memory and computational limitations when dealing with dense or highly populated jets. To overcome this structural bottleneck, architectures such as the IAFormer have proposed localized physics-informed architectures \cite{Esmail:2025kii}. Even with these architectures, it remains a challenge to scale up such dense interaction modules, which require a more delicate approach to balance the global context and local geometric precision. To overcome these limitations, we propose the \textit{S}parse \textit{Par}ticle-interaction \textit{T}ransformer \textit{A}rchitecture (\textbf{SParTA}), an architecture that factorizes the jet constituents into different topological regimes to circumvent the memory limitations of conventional attention. The network exploits a clear separation of hard (high-$p_T$) and soft (low-$p_T$) particles. This substantially reduces the quadratic memory footprint of the model, since local, high-fidelity interaction graphs are only constructed around the leading hard particles. Meanwhile, a soft-particle cross-attention mechanism retains global jet information so that low-energy radiation, which carries significant structural and color-flow information, is not lost but is dynamically aggregated back into the primary representations. The pairwise interaction tensor is computed dynamically during the forward pass from constituent 4-vectors. Jet-level global features are injected directly into the pairwise embeddings. Consequently, the learned interaction for any pair $(i, j)$ depends on the overall context of the entire jet, meaning the same pair of constituents will be embedded differently depending on the overall environment of the jet.

Furthermore, the SParTA enhances the relational representation via rich structural priors. The network employs explicit $p_T$-ordered Positional Encodings, as opposed to sequential coordinates, to directly embed the momentum hierarchy into the feature space. Within the localised attention blocks, the network employs a custom Differential Attention mechanism \cite{Esmail:2025kii}. The model is trained to learn the subtraction of background noise and the extraction of sharp structural boundaries by contrasting dual-head edge projections obtained from combined embedding and kinematic Lund features. This hybrid formulation enables the SParTA to maintain the expressive power of interaction-aware Transformers while scaling efficiently to high-multiplicity jet configurations.  

\begin{figure*}
    \centering
    \begin{subfigure}[t]{\textwidth}
        \centering
        \includegraphics[width=\textwidth]{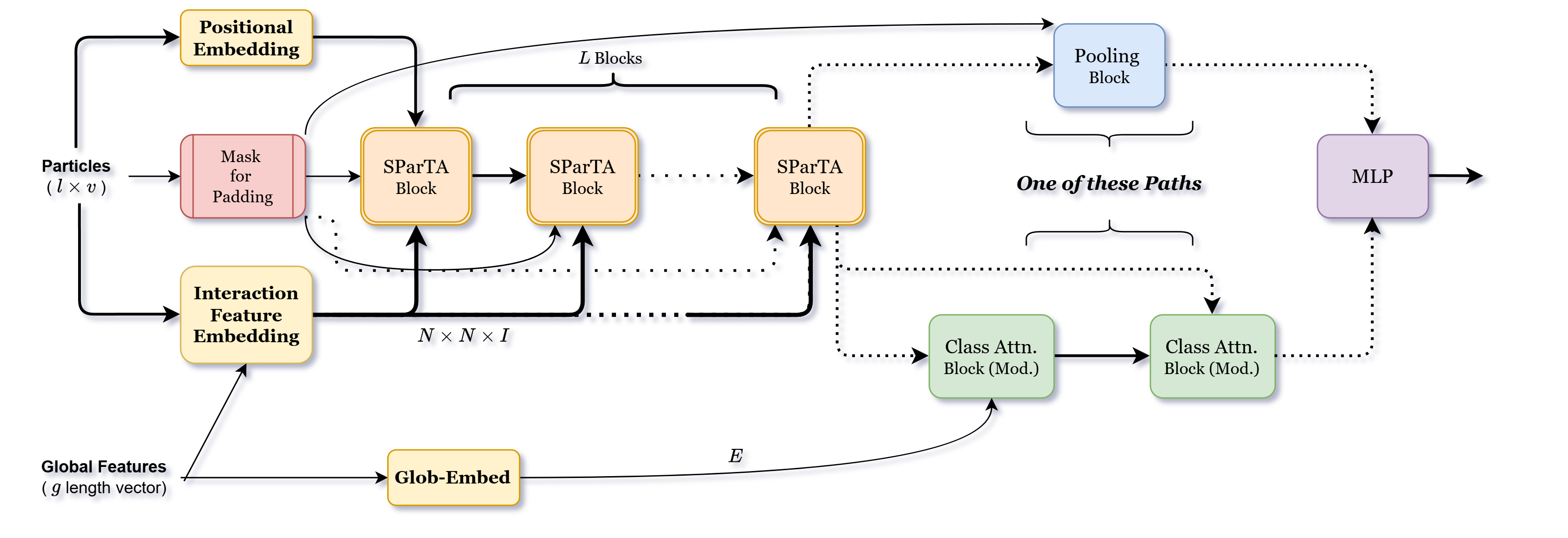}
        \caption{\small Full Schema}
        \label{fig:schema1_full}
    \end{subfigure}\\
    \begin{subfigure}[t]{\textwidth}
        \centering
        \includegraphics[width=0.95\textwidth]{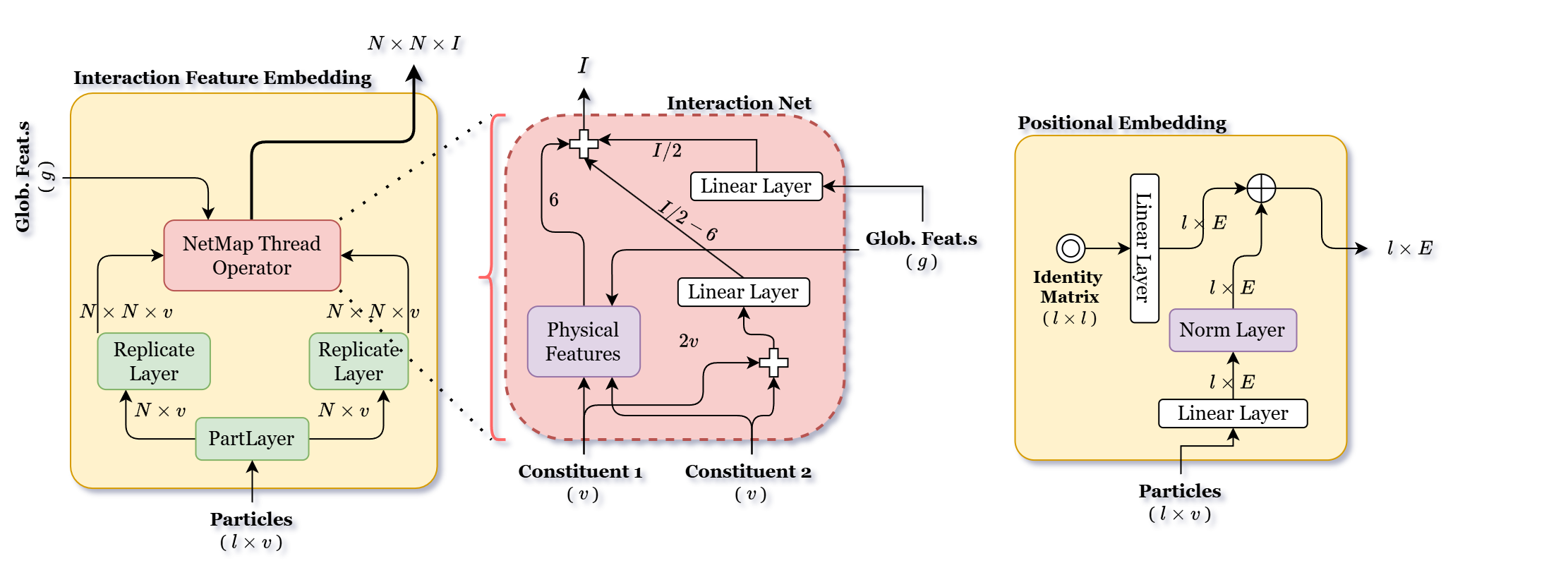}
        \caption{\small Embeddings}
        \label{fig:schema1_embed}
    \end{subfigure}\\
    \begin{subfigure}[t]{\textwidth}
        \centering
        \includegraphics[width=0.8\textwidth]{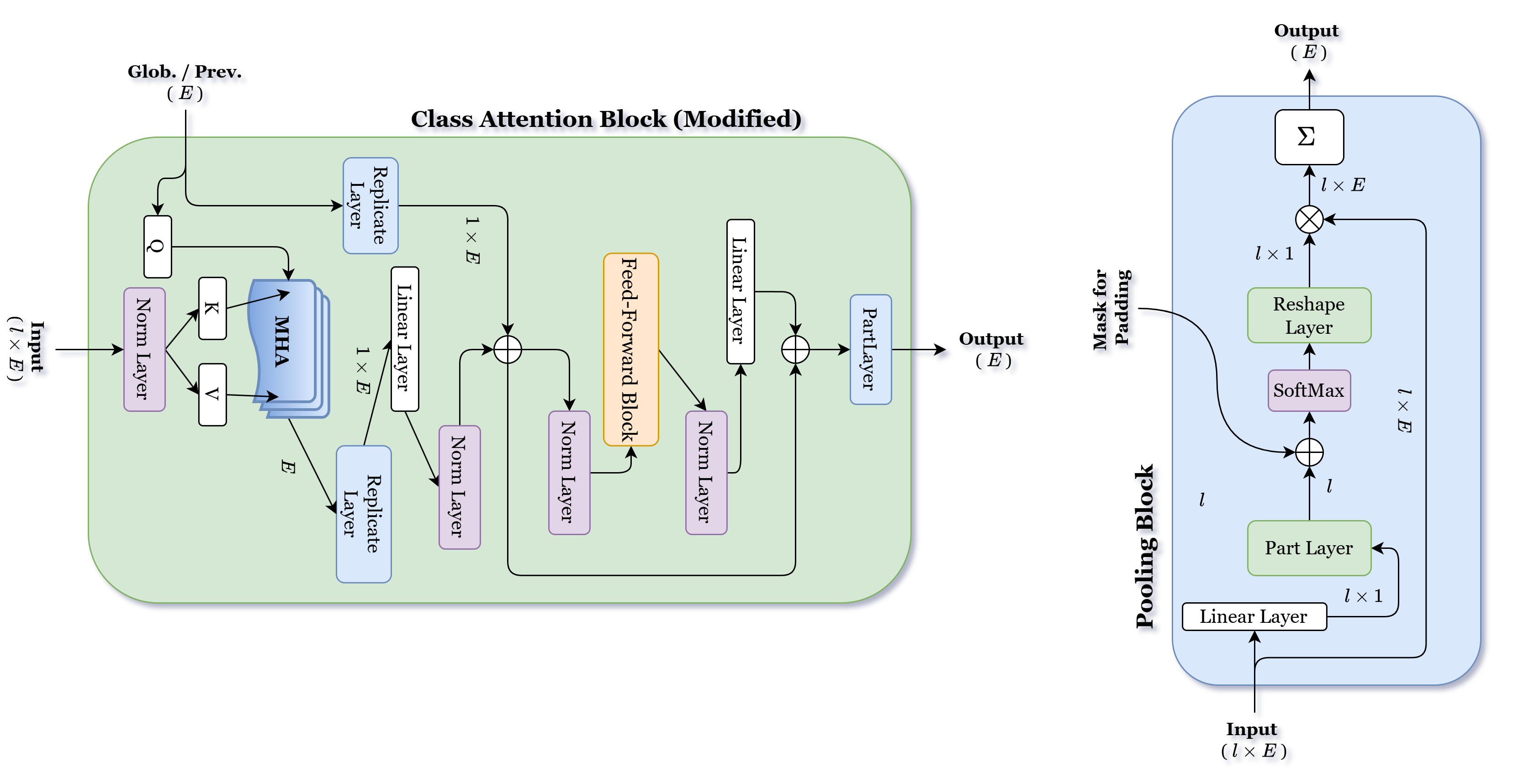}
        \caption{\small Classifier Blocks}
        \label{fig:schema1_last}
    \end{subfigure}
    \caption{\small Schematics of the SParT Architecture. }
    \label{fig:schema_1}
\end{figure*}

\section{Overview and data representation}
\label{sec:overview}
A jet is an unordered, variable-length set of reconstructed constituents. Following the standard \textit{particle-cloud} view~\citep{Qu:2019gqs, Qu:2022mxj}, each constituent is described by a vector of jet-relative kinematic features, and each jet additionally carries a vector of jet-level \emph{global} features. In SParTA the constituents are sorted by descending $\pt$ and zero-padded (or truncated) to a fixed length $l = 140$, so that the leading entries are always the hardest particles and the first $N$ of them form the hard core used by the interaction attention (Section~\ref{sec:ia}).
 
The model produces a single jet-level prediction from three ingredients that recur throughout: \emph{node} features (the sequence of per-constituent embeddings), \emph{edge} features (the pairwise interaction tensor), and \emph{global} features (the jet-level vector of high-level features, HLFs). What distinguishes SParTA from both plain transformers and GNNs is \emph{how} these three are combined, which is the subject of Sections~\ref{sec:architecture}. The rest of this section describes the two datasets and the node and global inputs built from them; the edge features are computed inside the network and are described in Section~\ref{sec:intEmbed}.
 
\subsection{Datasets}
\label{sec:datasets}
Our main benchmark is quark--gluon (QG) discrimination, on which the architecture scan of Section~\ref{sec:scan} is carried out. We also apply SParTA to top tagging, which tests the same architecture on a multi-prong, detector-level task. Both are public reference datasets used by the networks we compare against, and we use their recommended training, validation and test splits. Their main properties are collected in Table~\ref{tab:datasets}.
 
\begin{table}[!htb]
  \centering
  \small
  \caption{\small Properties of the two datasets and of the inputs built from them.}
  \label{tab:datasets}
  \setlength{\tabcolsep}{4pt}
  \begin{tabular}{@{}lcc@{}}
    \toprule
    & Quark--gluon & Top tagging \\
    \midrule
    Signal / background & $q$ / $g$ & $t$ / QCD \\
    Generator & \textsc{Pythia}~8 & \textsc{Pythia}~8 \\
    Detector sim. & none & \textsc{Delphes} (ATLAS) \\
    Jet radius $R$ & 0.4 & 0.8 \\
    Jet $\pt$ (GeV) & 500--550 & 550--650 \\
    Rapidity cut & $|y| < 1.7$ & $|\eta| < 2$ \\
    Train/val./test & 1.6M/0.2M/0.2M & 1.2M/0.4M/0.4M \\
    \midrule
    Padded length $l$ & 140 & 140 \\
    Features per constituent & 9 & 8 \\
    Global features & 53 & 53 \\
    \bottomrule
  \end{tabular}
\end{table}
 
\paragraph{Quark--gluon dataset.}
The QG sample of Komiske, Metodiev and Thaler~\citep{Komiske:2018cqr, Komiske:2019qgdata} is generated with \textsc{Pythia}~8 at $\sqrt{s} = 14$\,TeV from the processes $q\bar q \to Z(\to\nu\bar\nu) + g$ and $qg \to Z(\to\nu\bar\nu) + (u,d,s)$, with hadronization and the underlying event included and no detector simulation. Final-state non-neutrino particles are clustered with the anti-$k_{\mathrm T}$ algorithm~\citep{Cacciari:2008gp} with $R = 0.4$, and jets with $\pt \in [500, 550]$\,GeV and $|y| < 1.7$ are kept. The dataset contains $2\times10^6$ jets, evenly divided between quark (signal) and gluon (background) jets, split into $1.6\times10^6$ for training, $2\times10^5$ for validation and $2\times10^5$ for testing. Each constituent is stored as $(\pt, y, \phi, \mathrm{PID})$, where PID is the particle's PDG code.
 
\paragraph{Top-tagging dataset.}
The top-tagging reference dataset of Kasieczka \textit{et al.}~\citep{Kasieczka:2019topdata, Kasieczka:2019dbj} consists of hadronically decaying top-quark jets (signal) and QCD jets (background), generated with \textsc{Pythia}~8 at $\sqrt{s} = 14$\,TeV and passed through the \textsc{Delphes} ATLAS detector card, without multiple parton interactions or pile-up. Jets are clustered from Delphes energy-flow objects with anti-$k_{\mathrm T}$, $R = 0.8$, and are required to have $\pt \in [550, 650]$\,GeV and $|\eta| < 2$. A top jet must lie within $\Delta R = 0.8$ of the parton-level top quark, with all three top decay partons also within $\Delta R = 0.8$ of the jet axis. Each jet stores the four-momenta of up to 200 leading constituents; no particle identification is available. The $2\times10^6$ jets are split into $1.2\times10^6$ training, $4\times10^5$ validation and $4\times10^5$ test jets.

As in Ref.~\citep{Baruah:2025vpe}, jets with fewer than four constituents (33 jets) are removed, since $U_3$ and $M_3$ require at least four, in both the datasets.

\subsection{Constituent (sequence) features}
\label{sec:seqfeats}
Each constituent is described by the $v$ features of Table~\ref{tab:seqfeats}: the angular offsets from the jet axis, the constituent mass, the logarithms of its transverse momentum and energy, the same two quantities normalized to the jet, and its angular distance from the jet axis. These are the jet-relative kinematics used by ParT and IAFormer~\citep{Qu:2022mxj, Esmail:2025kii}. For the QG dataset a ninth feature encodes the particle identity. The $K$ distinct PDG codes present in the dataset are sorted in increasing order, $c_1 < c_2 < \dots < c_K$, and mapped onto equally spaced values in $[-1, 1]$,
\begin{equation}
  \mathrm{PID}(c_k) \;=\; -1 + \frac{2\,(k-1)}{K-1}, \qquad k = 1, \dots, K .
  \label{eq:pid}
\end{equation}
This single scalar replaces the one-hot particle-type and charge vector used by ParT and IAFormer~\citep{Qu:2022mxj, Esmail:2025kii}, keeping $v$ small. The top-tagging dataset has no particle identification, so $v = 8$ there.
 
\begin{table}[!htb]
  \centering
  \small
  \caption{\small Per-constituent input features. The PID feature is available for the QG dataset only ($v = 9$); top tagging uses the first eight ($v = 8$).}
  \label{tab:seqfeats}
  \setlength{\tabcolsep}{3pt}
\begin{tabular}{@{}cl>{\raggedright\arraybackslash}p{0.5\columnwidth}@{}}
    \toprule
    & Feature & Description \\
    \midrule
    1 & $\Delta\eta = \eta - \eta_{\mathrm{jet}}$ & pseudorapidity offset from the jet axis \\
    2 & $\Delta\phi = \phi - \phi_{\mathrm{jet}}$ & azimuthal offset from the jet axis \\
    3 & $m$ & constituent mass \\
    4 & $\log \pt$ & transverse momentum \\
    5 & $\log E$ & energy \\
    6 & $\log (\pt / \pt^{\mathrm{jet}})$ & momentum fraction \\
    7 & $\log (E / E^{\mathrm{jet}})$ & energy fraction \\
    8 & $\dR = \sqrt{\Delta\eta^2 + \Delta\phi^2}$ & distance from the jet axis \\
    9 & PID & particle identity, Eq.~\eqref{eq:pid} (QG only) \\
    \bottomrule
  \end{tabular}
\end{table}

The constituents of each jet are sorted by descending $\pt$ and the $l = 140$ leading ones are kept, with zero padding for jets of lower multiplicity; a padding mask derived from the all-zero rows excludes the padded positions from every attention and pooling step. The node features feed the token embedding of Section~\ref{sec:posEmbed}. Four of them, $\Delta\eta$, $\Delta\phi$, $\log\pt$ and $\log E$, combined with the jet axis from the global vector, also supply the absolute kinematics from which the pairwise interaction features of Eq.~\eqref{eq:lund} are computed inside the network (Section~\ref{sec:intEmbed}).
 
\subsection{Global (jet-level) features}
\label{sec:globfeats}
Each jet also carries a vector of 53 high-level features, the same for both datasets: its four-momentum, its constituent multiplicity, and a set of jet-substructure observables built from N-subjettiness and energy correlation functions (Table~\ref{tab:globfeats}). All are computed from the constituent four-momenta with \textsc{FastJet}~\citep{Cacciari:2011ma} and its \texttt{EnergyCorrelator} contrib, as in our earlier jet-image study~\citep{Baruah:2025vpe}, which used the same observables with a smaller set of N-subjettiness variables. Each feature is standardised to zero mean and unit variance before use.
 
\begin{table}[!htb]
  \centering
  \small
  \caption{\small The 53 jet-level (global) features used for both datasets. $\beta$ is the angular exponent; $(\alpha,\beta)$ label the two exponents of $D_2^{(\alpha,\beta)}$.}
  \label{tab:globfeats}
  \setlength{\tabcolsep}{4pt}
  \begin{tabular}{@{}l p{0.47\columnwidth} r@{}}
    \toprule
    Group & Features & No. \\
    \midrule
    Jet kinematics & $\pt^{\mathrm{jet}}$, $\eta_{\mathrm{jet}}$, $\phi_{\mathrm{jet}}$, $m_{\mathrm{jet}}$ & 4 \\
    Multiplicity & $N_{\mathrm{cons}}$ & 1 \\
    N-subjettiness~\citep{Thaler:2010tr, Thaler:2011gf} & $\tau_{21}^{\beta}$, $\tau_{32}^{\beta}$, $\tau_{43}^{\beta}$, $\tau_{54}^{\beta}$, $\tau_{65}^{\beta}$ for $\beta = 0.5$, $1.0$, $2.0$ & 15 \\
    $C$-series~\citep{Larkoski:2013eya} & $C_1, C_2, C_3$; $\beta = 1, 2$ & 6 \\
    $D$-series~\citep{Larkoski:2014gra, Larkoski:2015kga, Larkoski:2014zma, Bhattacherjee:2022gjq} & $D_2^{(\alpha,\beta)}$, $\alpha, \beta \in \{1, 2\}$; four $D_3$ variants & 8 \\
    $U$-series~\citep{Moult:2016cvt} & $U_1, U_2, U_3$; $\beta = 0.5, 1, 2$ & 9 \\
    $M$-series~\citep{Moult:2016cvt} & $M_2, M_3$; $\beta = 1, 2$ & 4 \\
    $N$-series~\citep{Moult:2016cvt} & $N_2, N_3$; $\beta = 1, 2$ & 4 \\
    $L$-series~\citep{ATL-PHYS-PUB-2021-028} & $L_2, L_3$ & 2 \\
    \midrule
    Total & & 53 \\
    \bottomrule
  \end{tabular}
\end{table}
 
The substructure groups probe complementary aspects of the radiation pattern. N-subjettiness ratios $\tau_{N,N-1}$ measure how well a jet is described by $N$ rather than $N-1$ subjets and are standard prong-counting observables for top tagging~\citep{Thaler:2010tr, Thaler:2011gf}. The energy-correlation ratios $C_N$ and $D_N$ are sensitive to the same prong structure without requiring subjet axes~\citep{Larkoski:2013eya, Larkoski:2014gra, Larkoski:2014zma}, while the generalized correlators $U$, $M$ and $N$ restrict the angular structure entering the correlation to target specific jet types, with the $U$-series aimed at quark--gluon discrimination~\citep{Moult:2016cvt}. Together with $N_{\mathrm{cons}}$ and the one-prong observables $C_1$ and $U_1$, these provide the classic analytic handles on quark--gluon separation.
 
The global vector enters SParTA in three places. The jet axis ($\eta_{\mathrm{jet}}$, $\phi_{\mathrm{jet}}$) and $\pt^{\mathrm{jet}}$ restore the absolute constituent kinematics needed for the pairwise features of Eq.~\eqref{eq:lund}. A learned projection of the full vector is concatenated to every edge, so that the embedded interaction of a pair $(i, j)$ depends on the state of the whole jet (Section~\ref{sec:intEmbed}). Finally, the vector is encoded and passed to the classification stage (Section~\ref{sec:head}). As discussed in Section~\ref{sec:architecture_feat}, a sufficiently expressive constituent-level network could in principle reconstruct these observables, but supplying them explicitly removes that burden~\citep{Mikuni:2020wpr, Baruah:2025vpe}.

\section{The architecture}
\label{sec:architecture}
As illustrated in Fig.~\ref{fig:schema1_full}, SParTA is organized into three modular stages: (i) an input encoding stage that computes $p_T$-ordered constituent token embeddings, a padding mask, an encoded global jet representation, and a dense $N \times N \times I$ pairwise interaction tensor for the $N$ hardest constituents; (ii) a stack of $B$ identical \emph{SParTA Blocks}, each coupling $O(N^2)$ sparse differential interaction attention on the hard core with $O(lN)$ cross-attention across all $l$ constituents; and (iii) a jet-level aggregation stage—using either attention pooling or a global-feature-driven class-attention block—followed by a multilayer perceptron (MLP) classification head.


\subsection{Input Embeddings}
\label{sec:embed}
\subsubsection{Constituent Token and $p_T$-Rank Positional Embedding}\label{sec:posEmbed}
Each constituent feature vector of length $v$ is mapped to the embedding dimension $E$ by a linear layer followed by RMS normalization~\citep{rmsnorm}. RMSNorm rescales by the root-mean-square of the activations without any bias; it is cheaper and empirically stabilises training in this setting, and it is the normalization adopted by IAFormer for the same reason~\citep{Esmail:2025kii}.

A learned positional encoding indexed by $p_T$ rank is then added to the token embeddings. Concretely, a fixed identity matrix is passed through a bias-free linear map, which is equivalent to a learnable embedding for each rank position; adding it makes the representation depend on a constituent's momentum ordering (see the `\textit{Positional Embedding}' block in fig. \ref{fig:schema1_embed}). This is a deliberate departure from the strict permutation-invariance that underlies Deep Sets and the Energy / Particle Flow Networks~\citep{deepsets, Komiske:2018cqr} and that ParT preserves~\citep{Qu:2022mxj}: we inject an ordering because $p_T$ rank is physically meaningful and the leading positions are precisely the ones the network will treat as the interaction core (Section~\ref{sec:ia}).

\subsubsection{Embedding the Interaction Tensor}\label{sec:intEmbed}
For the leading $N$ hard constituents, the network forms all ordered pairs $(i,j)$ and computes an interaction (edge) feature for each (see the `\textit{Interaction Feature Embedding}' block in fig. \ref{fig:schema1_embed}). 

Whereas six boost-related pairwise quantities are created for each of these pairs (following IAFormer~\citep[Eq.~11]{Esmail:2025kii}: 
\begin{equation}
\label{eq:lund}
I_{ij} = \Big[\,
\log\tfrac{\pt^i+\pt^j}{\pt^{\mathrm{jet}}},\;
\log\tfrac{E^i+E^j}{E^{\mathrm{jet}}},\;
\log \Delta_{ij},\;
\log k_{\mathrm{T}},\;
\log z,\;
\log m^2_{ij}
\,\Big]\, ,
\end{equation}
namely, the pair's summed transverse momentum and energy normalized to the jet, the angular distance, 
$$\Delta_{ij}=\sqrt{(\eta_i-\eta_j)^2+(\phi_i-\phi_j)^2} \, ,$$
the transverse-momentum scale $k_{\mathrm{T}}=\min(\pt^i,\pt^j)\,\Delta_{ij}$, the momentum sharing fraction $z=\min(\pt^i,\pt^j)/(\pt^i+\pt^j)$, and the squared invariant mass of the pair, all taken in logarithms.

The variables $z$, $k_{\mathrm{T}}$ and $\Delta$ are the coordinates of the primary Lund jet plane~\citep{lundplane}, in which successive QCD emissions populate well-separated kinematic regions; the summed momentum entries identify the most energetic pairs, which tend to form the hard prongs of a multi-pronged jet ~\citep{Esmail:2025kii}. The four core variables ($\Delta$, $k_{\mathrm{T}}$, $z$, $m^2$) are the interaction features introduced for jet tagging by ParT~\citep{Qu:2022mxj}. Crucially, the absolute kinematics needed here ($\eta$, $\phi$, jet energy) are reconstructed from the jet-level HLFs and the jet-relative per-particle node features; this is the first place the global vector enters the model. This part is depicted as the `\textit{Physical Features}' component of the interaction embedding block in fig. \ref{fig:schema1_embed}.

SParTA supports learnable interaction features. In the present implementation, an ordered pair is concatenated with a projection of the HLF vector (global features), so that the jet-level state modifies every edge. After concatenating the physical features to this, the resulting edge feature vector has length $I$. An optional convolution over the $N\times N$ interaction \textit{tensor} can be used to mix neighboring pairs before the attention stage. The interaction tensor is computed once from the inputs and made available to every block.


\subsection{SParTA Block}
\label{sec:spartaBlock}
\begin{figure*}
    \centering
    \includegraphics[width=0.95\textwidth]{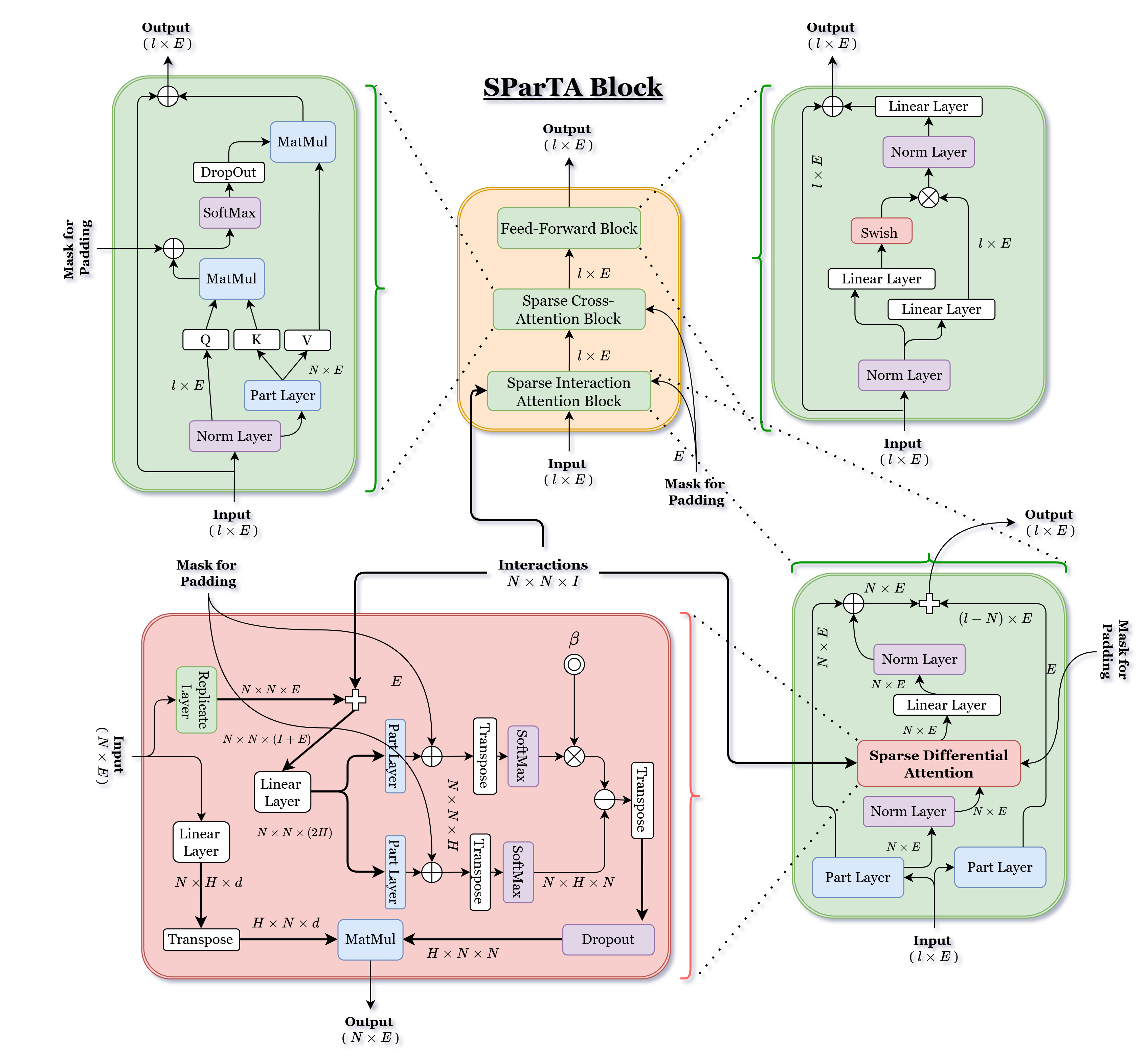}
    \caption{\small Schematics of the SParTA Block.}
    \label{fig:schema_2}
\end{figure*}
Each SParTA block takes two types of inputs: one $l\times E$ rectangular array and the $N\times N\times I$ interaction tensor. The rest of the subsection and Fig. \ref{fig:schema_2} explain the structure and functions of the three sub-blocks it has.

\subsubsection{Sparse Interaction Attention Block}\label{sec:ia}

On top of this, the block uses \emph{differential attention}~\citep{difftransformer}. Two separate projections $W_1,W_2$ of the same interaction matrix are each passed through a softmax, and the second is subtracted from the first:
\begin{equation}
\label{eq:diff}
\alpha \;=\; \softmax(W_1 \!\cdot\! I)\;-\;\beta\,\softmax(W_2 \!\cdot\! I),
\end{equation}
where $\beta$ is a learnable scalar, shared across heads within a block and clipped to $[0,1]$, exactly as in IAFormer~\citep[Eq.~9]{Esmail:2025kii}. The subtraction cancels the ``common-mode'' part of the two attention maps, suppressing diffuse weight on uninformative (typically soft) pairs and sharpening attention onto the few pairs that matter; in the original language-modelling setting this was motivated by analogy with a differential amplifier~\citep{difftransformer}. A consequence worth stating plainly is that $\alpha$ can be \emph{negative}, i.e., the effective adjacency is a signed, affine reweighting of the values rather than a convex average. Before the softmax, a diagonal mask removes self-interactions and the padded positions are masked out so that \textit{non-constituents} do not attract weight. This is represented throughout the schematics (Fig. \ref{fig:schema_1} and \ref{fig:schema_2}) as `\textit{Mask for Padding}' block and connections.

The computation is deliberately confined to the $N$ hard particles, so its cost is $O(N^2)$ rather than $O(l^2)$ in the full multiplicity $l$. This is the memory-driven design decision that gives the model its name. The full pairwise interaction is the expensive part, and we pay for it only through the hard constituents, where the discriminating substructure (prongs, hard splittings) lives. The interaction-attention output is added back to the hard tokens through a residual connection with RMSNorm, and the unchanged soft tokens are spliced back on.

\subsubsection{Cross-attention Block}\label{sec:cross}
The soft constituents still carry information (e.g., soft radiation patterns differ between quark- and gluon-initiated jets), so they must be coupled to the hard constituents. This is done with an ordinary multi-head dot-product cross-attention in which \emph{all} constituents form queries while the hard constituents supply keys and values. Its cost is $O(l N)$, linear in the multiplicity.

Structurally, this is the \textit{ideologically} similar to the construction of the Set Transformer~\citep{settransformer}, where a small set of representatives first absorbs information from the full set (here, the interaction-attention of Section~\ref{sec:ia} acts on the hard constituents), and the full set then attends back to those representatives, reducing a quadratic all-to-all interaction to two linear passes through a bottleneck. The difference is that our \textit{inducing-set} is not a set of free learned vectors but is chosen physically as the hardest particles, which makes the $p_T$ ordering of Section ~\ref{sec:embed} important.


\subsubsection{Feed-forward network}\label{sec:ffn}
Each block closes with a gated feed-forward network: a SwiGLU unit~\citep{gluvariants}, i.e. the element-wise product of a Swish / SiLU-activated~\citep{silu} linear branch with a plain linear branch, wrapped in RMSNorm and a residual connection in the usual pre-norm transformer arrangement. SwiGLU and the SiLU/GELU family of smooth activations~\citep{gluvariants, silu, gelu}
are the same components IAFormer uses for its per-token MLP~\citep{Esmail:2025kii}.

\subsection{Pooling and classification head}
\label{sec:head}
After the block stack, the variable-length set of token embeddings is reduced to a single vector by attention pooling: a learned score per token, a softmax over the sequence, and a weighted sum.
We use pooling rather than a dedicated class token; IAFormer makes the same choice and notes that, for particle clouds, replacing the class-attention aggregation of ParT/CaiT~\citep{Qu:2022mxj, touvron2021going} with
a simple pooling step is sufficient and cheaper~\citep{Esmail:2025kii}. The pooled vector is passed through a small normalized MLP to produce the class scores.

\section{Definition of the architectural features}
\label{sec:architecture_feat}

Table~\ref{tab:jet_tagging_feature_comparison} in Appendix~\ref{app:features} compares SParTA with seven other jet-tagging networks, feature by feature. This section defines those features and places SParTA among them.

\subsection{Symmetry and representation}
Except L-GATr \cite{Brehmer:2024yqw}, no other transformer-based architecture follows either the hard structural constraint of exact Lorentz equivariance. As a result, neither does every token carry scalar, vector, bivector, trivector, and pseudoscalar components, nor are the preferred beam axis and the preferred rest frame required. 
The only respected symmetries in most of these transformer networks are permutation invariance of the constituents (since the jet is physically an unordered set of particles) and longitudinal boost invariance of the pairwise inputs. While the latter is respected for the inputs in SParTA, the former is violated in favor of the $p_{\rm T}$ ordering. We 

\subsection{Pairwise interactions}
The most widely adopted way of injecting pairwise physics into a transformer, as adopted by ParT \cite{Qu:2022mxj} and MIParT \cite{Wu:2024thh}, is to add the embedded interaction tensor to the attention logits before the softmax (an additive attention bias):
\begin{equation}
\mathrm{Attn}(Q,K,V) = \mathrm{softmax}\!\left(\frac{QK^{\top}}{\sqrt{d}} + \mathcal{U}\right) V ,
\end{equation}
so that physics-motivated pair information modulates, but does not replace, the learned content-based attention. SParTA adds the interaction tensor before the attention mechanism, which is different from ParT or MIParT (as in IAFormer \cite{Esmail:2025kii}), where the query-key product is entirely removed and the attention map is obtained from the interaction tensor alone, e.g.,
\begin{equation}\label{eq:alpha_def}
\alpha = \mathrm{softmax}\!\left(W_1 \mathcal{U}\right) - \beta \, \mathrm{softmax}\!\left(W_2 \mathcal{U}\right),
\end{equation}
with only the value projection retained. Attention weights become a function of particle pairs rather than particle \emph{content}, which removes the $Q$ and $K$ parameter matrices and changes what attention can express. 

In MIParT, IAFormer, and SParTA, the interaction representation is re-embedded or re-projected at each block or independently per attention head, rather than being embedded once and reused unchanged throughout the network (a single embedded $\mathcal{U}$ is added identically at every layer in the baseline design). Re-learning the interactions at each block removes that restriction at the cost of additional parameters.

In case of pairwise quantities, we have only two options: either the tensors are precomputed and stored, fixing the features during pre-processing (this also multiplies the on-disk size of the dataset by $O(N^2)$ per event), or are computed inside the forward pass from the constituent four-vectors, so the network requires only per-particle kinematics as input. In SParTA, we chose the latter one. Pointwise ($1{\times}1$) convolutions can also be applied to the interaction tensor, so that each particle pair is transformed independently without mixing the features from different pairs. Though this facility is available in the SParTA blocks, we did not use it for the present set of results. The efficacy of these should be tested on larger datasets later.

When it comes to the pairwise embedding, SParTA is unique in two ways. First, in addition to the fixed, physically motivated features, a set of \textit{learned pairwise features} is produced by a learned map from the constituent pair. The fixed channels encode known QCD structure; the learned features allow the network to construct pairwise quantities that were not specified in advance. When concatenated and given equal weight, both are available to the attention mechanism. When just the physical features are embedded through an MLP, every output channel is still a function of the same handcrafted inputs. Second, jet-level quantities are also supplied as inputs to the pairwise embedding, so that the learned interaction for a pair $(i,j)$ depends on the properties of the whole jet, as well as the two constituents. This makes the pairwise representation context-dependent: the same pair of constituents can be embedded differently in jets of different mass or $p_T$.

\subsection{Attention and aggregation}
Similar to most earlier transformers---such as ParT, MIParT, and L-GATr---the primary mixing operation in SParTA is scaled dot-product attention, in which each token attends to all others through a learned query-key inner product. Following IAFormer, two separate attention maps are computed; then one is subtracted from the other, after scaling with a \textit{learned} $\beta$ factor (see eq. \ref{eq:alpha_def}). Since the softmax is strictly positive, a single attention map necessarily assigns non-zero weight to every token and hence, the difference of two maps can potentially cancel to zero or become negative, allowing the mechanism to actively suppress uninformative pairs rather than merely down-weighting them.

SParTA uniquely splits the processing of jet constituents. A leading subset of $N$ constituents undergoes full attention computation. The remaining constituents are processed using computationally cheaper methods. This restricts the quadratic cost to the hard subset, but the soft constituents still influence the final representation. This computational choice leverages the steeply falling $p_{\rm T}$ spectrum of jet constituents. While it introduces a new hyperparameter, $N$, the network effectively links fine-grained objects to a coarser summary. It achieves this by drawing queries from one token set and keys and values from another (this specific type of usage can be found in the Mixer-Net \cite{Hammad:2024cae} ). Ultimately, this approach reduces computational costs when the auxiliary set is small. 

Similar to ParT and MIParT, SParTA has the option to incorporate class-attention in the final blocks, for stabilization and accuracy boost (initially inspired from the \textit{CaiT} architecture \cite{touvron2021going}). In the original implementation, a dedicated learned token attends to the particle representations in one or more final blocks, and its output is passed to the classifier. Because the aggregation weights are learned and input-dependent, the network can emphasize the constituents most relevant to the decision, rather than weighting all of them equally. In the present implementation, the jet-level features are fed in through an encoding linear layer in place of the class tokens. SParTA also lets the user add a pooling-based aggregation in place of the class-attention mechanism. In that case, the constituent particle representations are reduced to a jet-level vector by a fixed \textit{symmetric} operation such as mean or max pooling. This is parameter-free and manifestly permutation invariant, but the aggregation weights cannot adapt to the input.

Another popular way of handling features in the literature is local neighborhood graph convolution (e.g., EdgeConv or kNN), where features are aggregated over the $k$ nearest neighbors of each constituent particle in some metric, typically $\eta - \phi$ space, using an edge function of the form $h(x_i, x_j - x_i)$. The inductive bias is locality: collinear splittings produce nearby constituents, so a local operator captures the substructure of a prong efficiently. Unlike attention, the neighborhood is determined geometrically rather than learned. Though it is possible to do this inside the `\textit{Interaction Feature Embedding}' block, we have not implemented it in the present iteration of SParTA. 

\subsection{Inputs, encoding and output}
It has been observed that scalar quantities describing the jet as a whole---known as global or high-level features---e.g., mass, $p_T$, multiplicity, and substructure observables, when explicitly supplied to the classifier network alongside the constituent-level features, improve performance \cite{Mikuni:2020wpr, Baruah:2025vpe}. In principle, a sufficiently expressive constituent-level network can reconstruct such quantities, but providing them directly removes that burden and can improve sample efficiency. 

SParTA has the provision to use these in different ways.

Another unique feature of SParTA is a learned embedding, indexed by the constituent's position in the (usually $p_T$-ordered) input list, which is added to the particle representation. This provides ordering information that a permutation-invariant network cannot otherwise access, at the cost of invariance itself; the trade-off is favorable only when the ordering carries genuine physical meaning.

The per-particle (node-level) output of ABCNet \cite{Mikuni:2020wpr}, where the network outputs a prediction for each constituent (useful for tasks such as pileup mitigation), is not implemented in SParTA.

The principal source of per-token nonlinearity in a transformer arises from the activation of the feed-forward sub-layer inside each attention block. The attention sub-layer is nonlinear in how it computes its weights, but acts linearly on the values, and stacked attention without a nonlinear feed-forward sub-layer degenerates towards rank-one representations with depth. SParTA uses a novel \emph{SwiGLU} activation, the gated variant $\big(W_1 x \odot \mathrm{Swish}(W_2 x)\big) W_3$, which uses three projections instead of two and multiplies one branch by a gate rather than applying a single element-wise function. These can be different from the activations appearing inside attention-coefficient computations.


\section{Analysis and performance}
\label{sec:analysis}


To understand which architectural choices drive SParTA's performance and at what cost, we trained 30 configurations on the QG dataset of Section~\ref{sec:datasets}. Performance is set mainly by depth up to four blocks and by the size $N$ of the hard core, while widening the network adds parameters without improving discrimination. The configuration used for the QG results of Table~\ref{tab:qgTagRes1}, 
$E_{60}H_{20}N_{80}F_{4}D_{2}B_{4}$ (300.4K parameters), ranks fifth in the scan with an AUC of 91.77\%, 0.09 percentage points below the best scan model and statistically indistinguishable from it at the precision of a single training run (Section~\ref{sec:scan-results}). This configuration alone was subsequently fine-tuned, which raised its AUC to 92.02\% with $1/\epsilon_B = 45.7$ and $121.8$ at $\epsilon_S = 0.5$ and $0.3$. All numbers in Sections~\ref{sec:scan}--\ref{sec:scan-cka} are for the scan models before fine-tuning.

\subsection{Search space and training}
\label{sec:scan}\label{sec:scan-procedure}
We ran a grid over four architectural hyperparameters and trained every point with one fixed recipe so that differences in performance can be attributed to the architecture alone. The number of attention heads was fixed at $H = 20$. The scanned values are listed in Table~\ref{tab:search-space}. The network files, which are compatible with both Mathematica and PyTorch are publicly available \href{https://github.com/BEC-HEP/SParTA}{here}.
 
\begin{table}[!htb]
  \centering
  \small
  \caption{\small Hyperparameters scanned. The number of attention heads is fixed at 20.}
  \label{tab:search-space}
  \begin{tabular}{@{}llc@{}}
    \toprule
    Hyperparameter & Symbol & Values scanned \\
    \midrule
    Embedding dimension & $E$ & 40, 60, 80 \\
    Hard-core size & $N$ & 60, 80 \\
    Feed-forward width & $F$ & $3E$, $4E$ \\
    Number of SParTA blocks & $B$ & 2, 4, 6 \\
    \bottomrule
  \end{tabular}
\end{table}
 
The grid is complete for $E = 40$ and $60$ ($2 \times 2 \times 3 = 12$ models each). For $E = 80$ only $N = 80$ was trained (6 models), giving 30 configurations in total. Dropout was set by model size rather than scanned: 0.2 for blocks with fewer than 90k trainable parameters (all $E = 40$ and $E = 60$ models) and 0.3 above that (all $E = 80$ models). A configuration is labeled $E_eH_hN_nF_fD_dB_b$; for example, $E_{40}H_{20}N_{80}F_{3}D_{2}B_6$ has an embedding dimension of $40$, $20$ heads, $80$ hard constituents, feed-forward width $120 ~[3\times 40]$, dropout $0.2$, and $6$ blocks.
 
The parameter count depends only on $E$, $F$ and $B$. The hard-core size $N$ changes the size of the interaction tensor and the attention maps, and hence the FLOPs, but adds no trainable weights. Configurations that differ only in $N$ therefore isolate the effect of how many constituents enter the interaction attention at fixed model capacity.
 
Every model was trained with Adam~\citep{kingma2014adam} ($\beta_1 = 0.9$, $\beta_2 = 0.999$, $\epsilon = 10^{-8}$), cross-entropy loss, gradient-norm clipping at 1.0 and a batch size of 512. The learning rate starts at $5 \times 10^{-3}$ and is halved whenever the validation loss fails to improve (relative change below $10^{-5}$) for more than 3 epochs, down to $10^{-6}$. Training stops after 12 epochs without improvement, or at 200 epochs, and the checkpoint with the lowest validation loss is kept. A fixed batch size was used for every model, rather than the largest that fits in GPU memory, so that the scan compares architectures and not effective learning rates. Each configuration was trained once, on a single NVIDIA A100 80\,GB GPU.
 
Each kept checkpoint was evaluated on the 199{,}856-jet test set. We report the AUC and the background rejection $1/\epsilon_B$ at $\epsilon_S = 0.5$ and $0.3$ ($R_{50}$, $R_{30}$), with quark jets as signal. FLOPs are counted for a single forward pass of one jet.
 
\subsection{Scan results}
\label{sec:scan-results}
All 30 configurations lie within 0.6 percentage points of AUC (91.27--91.86\%), and the top ten lie within 0.17 points of each other (Table~\ref{tab:scan-results}). The test set itself limits how finely they can be ranked. With about $10^5$ jets per class, the statistical uncertainty is roughly 0.065 percentage points on the AUC~\citep{Hanley:1982auc}, about $\pm 1$ on $R_{50}$ and $\pm 4$--5 on $R_{30}$ (binomial on the number of background jets passing the cut). Seed-to-seed variation, not measured here, adds to this.
 
\begin{table*}[!tb]
  \centering
  \small
  \setlength{\tabcolsep}{4pt}
  \caption{\small All 30 scanned SParTA configurations on the QG test set, ranked by AUC, before fine-tuning. $B$: blocks; $E$: embedding dimension; $N$: hard-core size; Dp: dropout; FFN: feed-forward width $F$. $R_{50}$ and $R_{30}$ are the background rejections $1/\epsilon_B$ at $\epsilon_S = 0.5$ and $0.3$. $^\star$Configuration used, after fine-tuning, for the results of Table~\ref{tab:qgTagRes1}.}
  \label{tab:scan-results}
  \begin{tabular}{@{}rrrrrrrrrrr@{}}
    \toprule
    Rank & $B$ & $E$ & $N$ & Dp & FFN & Params (k) & FLOPs (M) & AUC (\%) & $R_{50}$ & $R_{30}$ \\
    \midrule
    1 & 6 & 40 & 80 & 0.2 & 120 & 185.3 & 416.1 & 91.86 & 46.2 & 126.5 \\
    2 & 6 & 80 & 80 & 0.3 & 320 & 757.3 & 695.8 & 91.80 & 44.8 & 124.1 \\
    3 & 6 & 40 & 80 & 0.2 & 160 & 214.9 & 424.4 & 91.79 & 45.1 & 126.3 \\
    4 & 4 & 40 & 80 & 0.2 & 160 & 146.9 & 293.1 & 91.78 & 45.4 & 121.4 \\
    5* & 4 & 60 & 80 & 0.3 & 240 & 300.4 & 376.5 & 91.77 & 45.2 & 123.1 \\
    6 & 4 & 80 & 80 & 0.3 & 320 & 511.5 & 474.6 & 91.77 & 44.9 & 126.5 \\
    7 & 4 & 80 & 80 & 0.3 & 240 & 433.7 & 452.8 & 91.75 & 44.4 & 123.8 \\
    8 & 4 & 40 & 80 & 0.2 & 120 & 127.3 & 287.5 & 91.72 & 45.0 & 121.3 \\
    9 & 6 & 60 & 80 & 0.2 & 240 & 442.9 & 549.2 & 91.72 & 44.4 & 121.0 \\
    10 & 4 & 40 & 60 & 0.2 & 120 & 127.3 & 174.3 & 91.69 & 45.6 & 123.1 \\
    11 & 6 & 40 & 60 & 0.2 & 160 & 214.9 & 260.8 & 91.67 & 45.4 & 123.1 \\
    12 & 4 & 60 & 80 & 0.2 & 180 & 256.5 & 364.2 & 91.66 & 44.2 & 117.3 \\
    13 & 6 & 60 & 60 & 0.2 & 240 & 442.9 & 354.7 & 91.65 & 44.9 & 124.3 \\
    14 & 2 & 40 & 80 & 0.2 & 120 & 69.2 & 158.9 & 91.64 & 43.9 & 122.6 \\
    15 & 4 & 40 & 60 & 0.2 & 160 & 146.9 & 179.8 & 91.63 & 45.6 & 123.4 \\
    16 & 6 & 40 & 60 & 0.2 & 120 & 185.3 & 252.5 & 91.63 & 45.0 & 119.7 \\
    17 & 6 & 60 & 80 & 0.2 & 180 & 377.0 & 530.7 & 91.59 & 43.1 & 118.3 \\
    18 & 6 & 60 & 60 & 0.2 & 180 & 377.0 & 336.2 & 91.58 & 44.4 & 119.5 \\
    19 & 2 & 40 & 80 & 0.2 & 160 & 79.0 & 161.7 & 91.56 & 43.2 & 115.9 \\
    20 & 4 & 60 & 60 & 0.2 & 240 & 300.4 & 242.7 & 91.56 & 45.1 & 118.4 \\
    21 & 2 & 40 & 60 & 0.2 & 120 & 69.2 & 96.0 & 91.55 & 43.9 & 120.0 \\
    22 & 2 & 80 & 80 & 0.3 & 320 & 265.6 & 253.3 & 91.55 & 43.8 & 118.8 \\
    23 & 4 & 60 & 60 & 0.2 & 180 & 256.5 & 230.4 & 91.54 & 44.1 & 119.7 \\
    24 & 2 & 60 & 80 & 0.2 & 180 & 136.0 & 197.7 & 91.53 & 43.5 & 117.2 \\
    25 & 2 & 60 & 80 & 0.2 & 240 & 157.9 & 203.9 & 91.53 & 43.3 & 117.0 \\
    26 & 2 & 40 & 60 & 0.2 & 160 & 79.0 & 98.8 & 91.52 & 44.2 & 119.0 \\
    27 & 6 & 80 & 80 & 0.2 & 240 & 640.7 & 663.1 & 91.47 & 41.7 & 110.2 \\
    28 & 2 & 60 & 60 & 0.2 & 240 & 157.9 & 130.7 & 91.38 & 43.0 & 115.1 \\
    29 & 2 & 80 & 80 & 0.3 & 240 & 226.7 & 242.4 & 91.32 & 42.1 & 107.7 \\
    30 & 2 & 60 & 60 & 0.2 & 180 & 136.0 & 124.5 & 91.27 & 42.6 & 109.6 \\
    \bottomrule
  \end{tabular}
\end{table*}
 
\subsection{Effect of individual hyperparameters}
\label{sec:scan-effects}
Because the scan is a grid, the effect of one hyperparameter can be measured directly: we compare every pair of configurations that differ in that hyperparameter only, with the other three held fixed, and record the change in AUC. Table~\ref{tab:paired} gives, for each step, the number of such pairs, the mean and range of the AUC change, and how many pairs improved. Individual differences are mostly within the test-set uncertainty, so we treat a trend as established only when its sign is the same across all pairs; for 12 out of 12 pairs this has a two-sided sign-test probability of $5 \times 10^{-4}$ under the null hypothesis of no effect.
 
\begin{table}[!htb]
  \centering
  \footnotesize
  \caption{\small Change in test AUC (percentage points) and in $R_{30}$ for one hyperparameter step, over all $n$ pairs of configurations that differ only in that step; ``Up'' counts the pairs whose AUC improved. $F$ is compared as a multiple of $E$; the $E\!: 60 \to 80$ pairs all have $N = 80$.}
  \label{tab:paired}
  \setlength{\tabcolsep}{3pt}
  \begin{tabular}{@{}lccccc@{}}
    \toprule
    Step & $n$ & $\langle\Delta\mathrm{AUC}\rangle$ & Range & Up & $\langle\Delta R_{30}\rangle$ \\
    \midrule
    $B\!: 2 \to 4$ & 10 & $+0.202$ & $[+0.08, +0.43]$ & 10 & $+5.5$ \\
    $N\!: 60 \to 80$ & 12 & $+0.123$ & $[+0.01, +0.26]$ & 12 & $+1.1$ \\
    $F\!: 3E \to 4E$ & 15 & $+0.059$ & $[-0.08, +0.33]$ & 10 & $+2.7$ \\
    $B\!: 4 \to 6$ & 10 & $-0.011$ & $[-0.28, +0.14]$ & 6 & $-0.5$ \\
    $E\!: 60 \to 80$ & 6 & $-0.023$ & $[-0.21, +0.09]$ & 3 & $-0.5$ \\
    $E\!: 40 \to 60$ & 12 & $-0.105$ & $[-0.28, -0.01]$ & 0 & $-3.5$ \\
    \bottomrule
  \end{tabular}
\end{table}

Two steps improve the AUC in every pair. Going from two to four blocks gives the largest gain, while a further step to six blocks gives none on average. This saturation agrees with IAFormer, which found quark--gluon performance to saturate with depth and halved its network from 12 to 6 layers for this task~\citep{Esmail:2025kii}. Enlarging the hard core from 60 to 80 constituents also helps in every pair, by 0.12 points on average, at no cost in parameters and 1.55--1.65 times the FLOPs. Since constituent multiplicity and the soft radiation pattern are among the strongest quark--gluon discriminants, treating more constituents with full pairwise interactions, rather than only through the hard--soft cross-attention, plausibly adds information.
 
Widening the network does not help. Raising $E$ from 40 to 60 lowers the AUC in all 12 pairs, and raising it from 60 to 80 changes it inconsistently. A $4E$ rather than $3E$ feed-forward width is neutral at $E = 40$ (mean $-0.02$, 2 of 6 pairs improved) and helps only the wider models ($+0.07$ at $E = 60$, $+0.19$ at $E = 80$). The two largest models, with 641k and 757k parameters, do not outperform the 185k-parameter model at the top of Table~\ref{tab:scan-results}. Accordingly, across the 30 models the AUC correlates only weakly with the parameter count (Spearman $\rho = 0.26$) but clearly with the FLOPs ($\rho = 0.61$): the two hyperparameters that matter, $B$ and $N$, both raise the FLOPs, while the one that adds most parameters, $E$, does not improve performance.
 
\subsection{Pareto efficiency}
\label{sec:scan-pareto}
\begin{figure*}[!tb]
  \centering
    \includegraphics[width=0.98\textwidth]{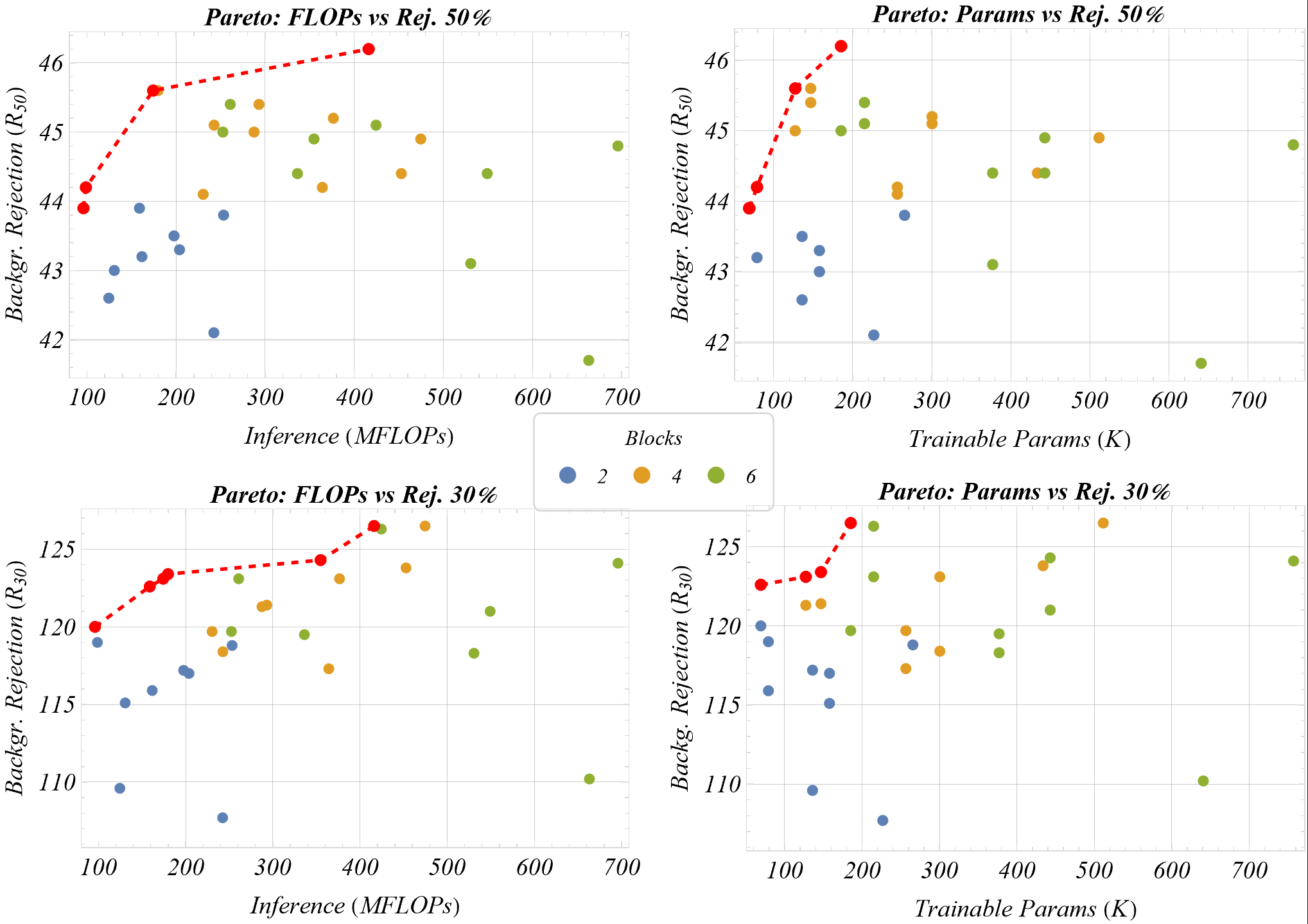}
  \caption{\small Background rejection against cost for the scanned SParTA configurations on the QG test set. Point colour indicates the number of blocks (blue: 2, green: 4, orange: 6); red points joined by the dashed line form the Pareto front.}
  \label{fig:pareto}
\end{figure*}

A tagger is useful only if its performance justifies its cost, so we compare the scanned configurations in the plane of a cost measure against a performance measure (Fig.~\ref{fig:pareto}). A configuration is \emph{Pareto-optimal}, or non-dominated, if no other configuration is at least as cheap and at the same time achieves a higher background rejection. The Pareto-optimal configurations form the Pareto front, which traces the best rejection attainable at each cost; every configuration below the front is dominated, meaning that some other configuration is both cheaper and better. We use two cost measures, the number of trainable parameters, which sets the memory footprint and model capacity, and the inference FLOPs per jet, which set the computational cost, and two working points, $R_{50}$ and $R_{30}$. The front is thus the natural place to choose a configuration for a given computational budget, and configurations far below it are inefficient uses of capacity or compute.
 

In all four planes the front is populated almost entirely by the narrowest models, $E = 40$; the only exception is the $E_{60}H_{20}N_{60}F_4D_2B_6$ on the FLOPs--$R_{30}$ front. This is the efficiency counterpart of Section~\ref{sec:scan-effects}: the extra width of the $E = 60$ and $E = 80$ models costs parameters and FLOPs without a matching gain in rejection.

\subsection{Variance decomposition}
\label{sec:scan-fanova}
A complementary view asks how much of the spread in a quantity $y$ across the scan each hyperparameter accounts for. For a hyperparameter $x_i$ we compute the main-effect variance fraction
\begin{equation}
  \eta_i^2 \;=\; \frac{\mathrm{Var}_{x_i}\!\big[\,\mathbb{E}(y \mid x_i)\,\big]}{\mathrm{Var}(y)} ,
  \label{eq:eta2}
\end{equation}
the share of the variance of $y$ explained by the mean of $y$ at each value of $x_i$. It is evaluated directly over the 30 trained configurations, with no surrogate model, treating the feed-forward width $F$ by its absolute value. On a complete, balanced factorial grid $\eta_i^2$ coincides with the first-order term of the functional ANOVA (fANOVA) decomposition used for hyperparameter importance~\citep{Hutter:2014fanova}, and the remainder $1 - \sum_i \eta_i^2$ is due to interactions. As $F\in\{3E, 4E\}$ and the $N = 60$ case is completely absent for $E = 80$, the 30-configuration scan is non-orthogonal (full scan should be of 36); thus, $\eta_i^2$ measures unadjusted marginal variance fractions (which overlap between $E$ and $F$) rather than an orthogonal decomposition summing to $1$.
 
\begin{figure}[!htb]
  \centering
  \includegraphics[width=0.8\textwidth]{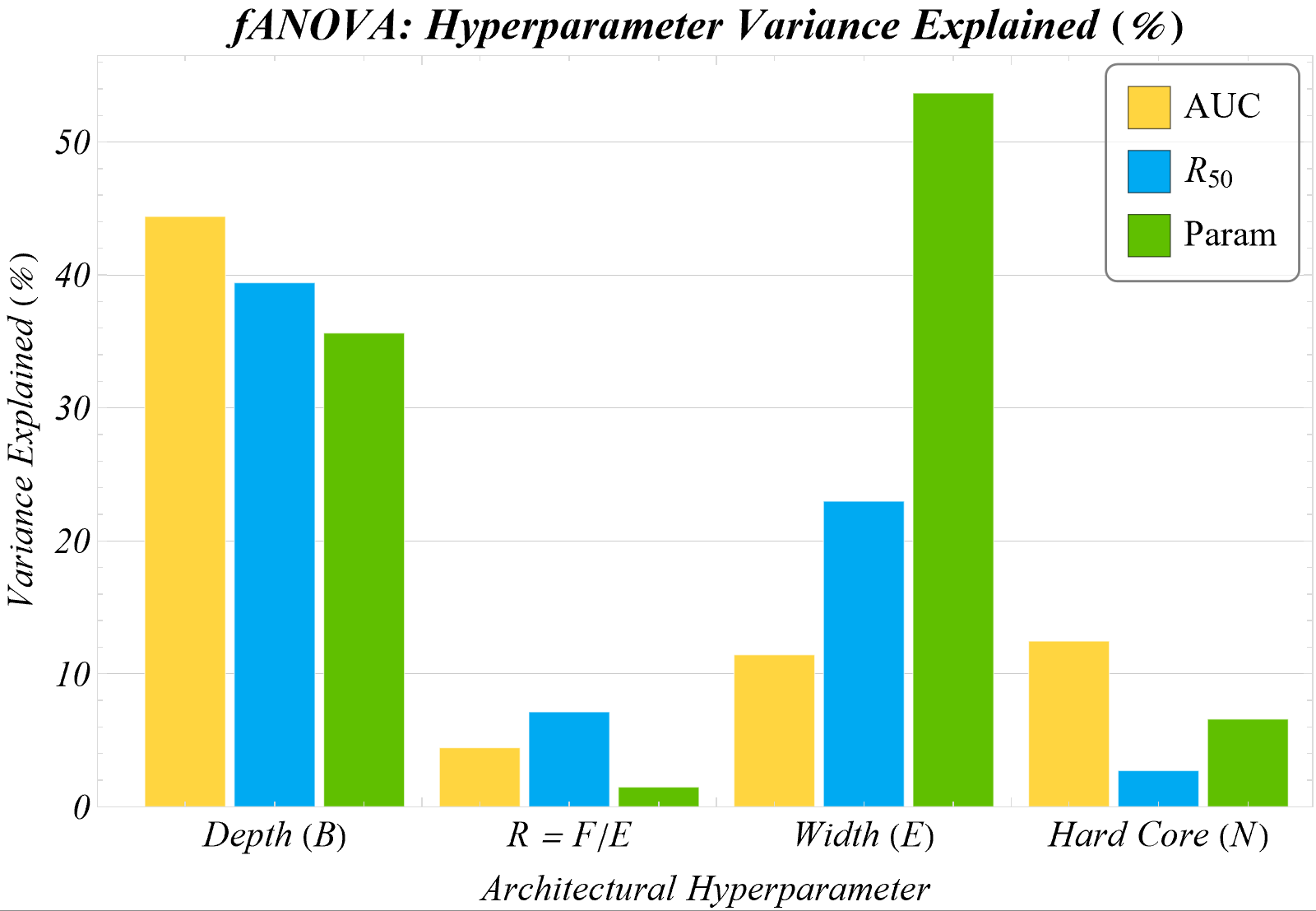}
  \caption{\small Main-effect variance fraction $\eta_i^2$, Eq.~\eqref{eq:eta2}, of each scanned hyperparameter for the test AUC (yellow), $R_{50}$ (blue) and the number of trainable parameters (green), over the 30 configurations of Table~\ref{tab:scan-results}.}
  \label{fig:fanova}
\end{figure}
 
Figure~\ref{fig:fanova} shows $\eta_i^2$ for the AUC, for $R_{50}$ and for the parameter count. For the AUC, depth ($B$) dominates with 44.4\% of the variance, followed by $N$ (12.5\%), $E$ (11\%), and $R$ ($F/E$, 4.4\%), leaving about 27\% unexplained, or resulting from interactions. 
 
The parameter count behaves differently: it is driven by $E$ (54\%) and $B$ (36\%), so the hyperparameters that set the cost of the model are not those that set its discrimination. This decoupling is why the Pareto fronts of Fig.~\ref{fig:pareto} are populated by narrow models. The small 6.6\% attributed to $N$, which adds no parameters, comes from the incomplete grid: $E = 80$ models exist only with $N = 80$, so $N$ partly stands in for $E$. On the balanced $E \in \{40, 60\}$ subgrid its share vanishes. 

\subsection{Similarity of block representations}
\label{sec:scan-cka}

To see how the representation evolves through the block stack, we compare the outputs of the four SParTA blocks of the fine-tuned $E_{40}H_{20}N_{80}F_3D_2B_6$ model using linear centered kernel alignment (CKA)~\citep{Kornblith:2019cka}, as done for IAFormer~\citep{Esmail:2025kii}. For two representations $X$ and $Y$ of the same $n$ jets, with Gram matrices $K = XX^{\top}$ and $L = YY^{\top}$,
\begin{align}
  \mathrm{CKA}(K, L) &= \frac{\mathrm{HSIC}(K, L)}{\sqrt{\mathrm{HSIC}(K, K)\,\mathrm{HSIC}(L, L)}}, \label{eq:cka}\\
  \mathrm{HSIC}(K, L) &= \frac{\mathrm{Tr}(KHLH)}{(n-1)^2}, \nonumber
\end{align}
where $H = I_n - \tfrac{1}{n}\mathbf{1}\mathbf{1}^{\top}$ is the centring matrix, $I_n$ the identity and $\mathbf{1}$ the vector of ones. CKA is invariant under orthogonal transformations and isotropic rescaling of the features, so it compares what two layers represent rather than how the features are arranged; it equals 1 for identical representations. Each block output is averaged over the valid constituents of each jet before the comparison. For the similarity calculation we used $n=5000$.
 
\begin{figure}[!htb]
  \centering
  \includegraphics[width=0.6\textwidth]{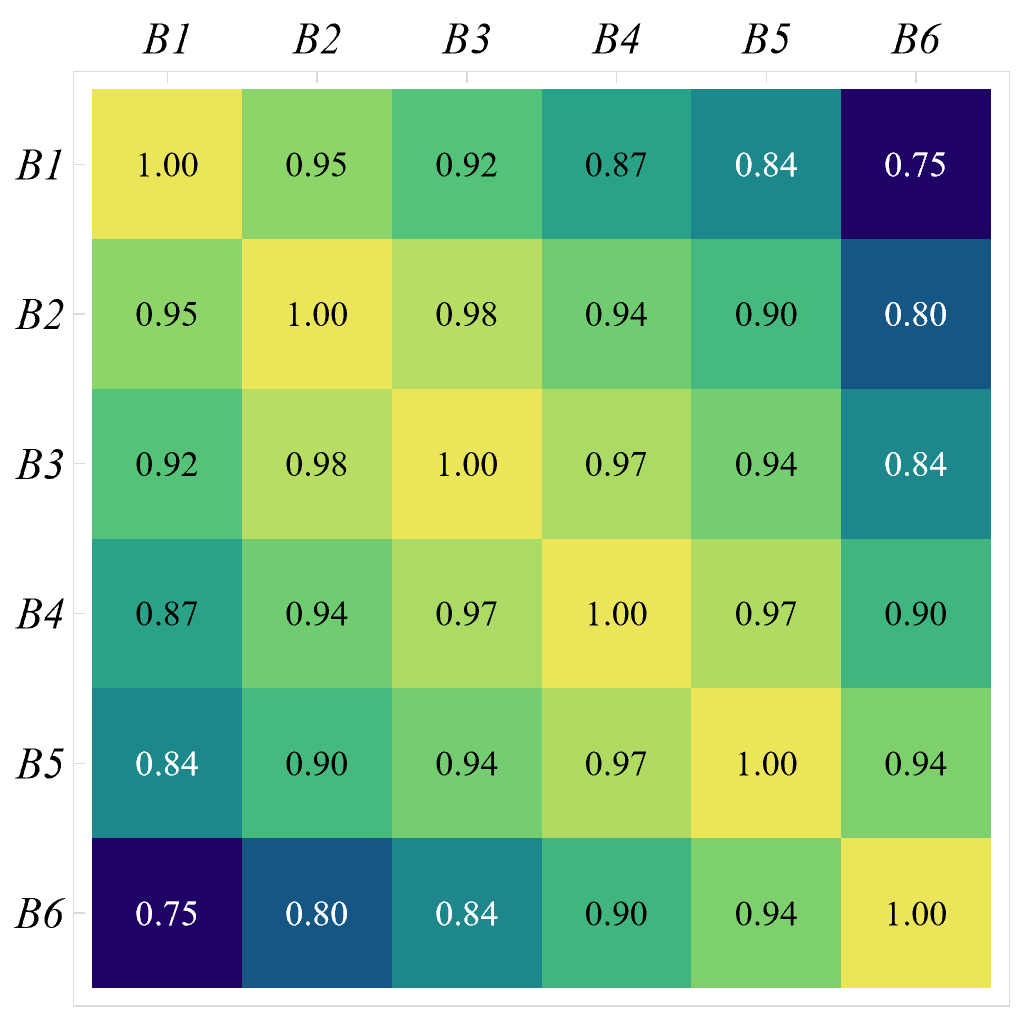}%
  \caption{\small Linear CKA, Eq.~\eqref{eq:cka}, between the outputs of the six blocks (B1--B6) of the fine-tuned $E_{60}H_{20}N_{80}F_4D_2B_4$ model on the QG test set.}
  \label{fig:cka}
\end{figure}
 

\subsection{Comparison with other taggers}
\label{sec:comparison}
Tables~\ref{tab:qgTagRes1} and~\ref{tab:topTagRes1} compare SParTA with published taggers on the two benchmarks. For SParTA we quote the best of $10$ independent trainings, and in the ``Dist.'' columns the mean and standard deviation over them; the numbers for all other networks are taken from their publications. Following Ref.~\citep{Esmail:2025kii}, the networks are grouped into Lorentz-invariant or -equivariant networks, attention-based networks trained from scratch on each benchmark, and networks pre-trained on the 100M-jet JetClass dataset~\citep{qu_2022_6619768} and then fine-tuned. On QG, SParTA is the $E_{60}H_{20}N_{80}F_4D_2B_4$ configuration of Section~\ref{sec:scan}. Its additional training at the end of Section~\ref{sec:analysis} used only the QG training set, so it belongs to the trained-from-scratch group.

On quark--gluon tagging, SParTA reaches an AUC of 92.02\% and an accuracy of 85.06\% with 300k parameters. Its best and mean AUC differ by only 0.01 percentage points, so the result does not depend on a favourable seed. The AUC matches that of ParT$_\text{full}$ (92.03\%), which has 2.14M parameters, about seven times more, and lies 0.13 points below MIParT$_\text{full}$ (92.15\%, 720.9k parameters). Against IAFormer, its closest relative, SParTA raises the AUC by 0.30 points and the background rejection from 42.0 to 45.7 at $\epsilon_S = 0.5$ and from 101.0 to 121.8 at $\epsilon_S = 0.3$, improvements of 9\% and 21\%. It also outperforms ParticleNet, LorentzNet, ABCNet, PCT and the pre-trained OmniLearn on every metric. At the two working points, however, ParT$_\text{full}$ and MIParT$_\text{full}$ still reject more background, and PELICAN remains the best network on every metric, with 208k parameters. Two differences in the inputs should be kept in mind. ParT$_\text{full}$ and MIParT$_\text{full}$ apply additional weights to charged and neutral hadrons~\citep{Qu:2022mxj}, while SParTA encodes particle identity in the single scalar of Eq.~\eqref{eq:pid}. Conversely, SParTA receives the 53 global features of Table~\ref{tab:globfeats}, which most of the compared networks do not use, so part of its gain may come from these inputs.

On top tagging, SParTA reaches an AUC of 98.60\% and background rejections of 443 and 1749 at $\epsilon_S = 0.5$ and $0.3$, with 300k parameters. It outperforms ParT (98.58\%, 413 and 1602) with a seventh of its parameters, and it exceeds ParticleNet and PCT at both working points and the Mixer network at $\epsilon_S = 0.5$. It falls short, however, of IAFormer (98.70\%, 510 and 2012), MIParT, and the Lorentz-equivariant LorentzNet, PELICAN and L-GATr; at $\epsilon_S = 0.3$ its rejection is about 13\% below that of IAFormer. 

\begin{table*}[!tb]
    \centering
    \small
    \renewcommand{\arraystretch}{1.25}
    \setlength{\tabcolsep}{4pt}
        \caption{\small Performance of jet taggers on the QG test set. Numbers for SParTA are from this work: ``Best'' is the best of [$10$] trainings and ``Dist.'' the mean and standard deviation over them. All other numbers are taken from the cited references. The subscript \textit{full} marks models that use additional weights for charged and neutral hadrons~\citep{Qu:2022mxj}, and f.t.\ marks models pre-trained on JetClass and fine-tuned. Bold marks the best value in each column and italics the best within each group.}
    \resizebox{\textwidth}{!}{%
    \begin{tabular}{llccccccc}
    \hline\hline
        & & \multicolumn{2}{c}{Accuracy (\%)}  & \multicolumn{2}{c}{AUC (\%)}   & \multicolumn{2}{c}{$1/\epsilon_B$} & \# of Params\\ 
        & & Best & Dist. & Best & Dist. & $\epsilon_S = 0.5$   & $\epsilon_S = 0.3$ & \\
        \hline
        & ParticleNet \cite{Qu:2019gqs} & $84.0$ & & $91.16$ & & $39.8(2)$ & $98.6(1.3)$ & 370K \\
        & PFN \cite{Komiske:2018cqr} & -- & & $90.52$ & & $37.4(7)$ & -- & \textbf{86.1K} \\
        & rPCN \cite{Shimmin:2021pkm} & -- & & $90.81$ & & $38.6(5)$ & -- & -- \\
        \hline
        \textit{Lorentz Invariance-} & PELICAN \cite{Bogatskiy:2023nnw} & $\mathbf{85.51}$ & & $\mathbf{92.52}$ & & $\mathbf{52.3(3)}$ & $\mathbf{149.8(2.4)}$ & 208K \\
        \textit{based} & LorentzNet \cite{Gong:2022lye} & $84.40$ & & $91.56$ & & $42.4(4)$ & $110.2(1.3)$ & 224K \\
        \hline\hline
        \textit{Attention-based} & ABCNet \cite{Mikuni:2020wpr} & $84.0$ & & $91.26$ & & $42.6(4)$ & $118.4(1.5)$ & 230K \\
        & PCT \cite{Mikuni:2021pou} & $84.1$ & & $91.40$ & & $43.2(7)$ & $118(2.2)$ & 193.3K \\
        \textbf{(\textit{Trained}} & ParT$_\text{full}$ \cite{Qu:2022mxj} & $84.9$ & & $92.03$ & & $47.9(5)$ & $129.5(9)$ & 2.14M \\
        \textbf{\textit{from}} & MIParT$_\text{full}$ \cite{Wu:2024thh} & $\textit{85.1}$ & & $\textit{92.15}$ & & $\textit{49.3(4)}$ & $\textit{133.9(1.4)}$ & 720.9K \\
        \textbf{\textit{scratch})} & IAFormer \cite{Esmail:2025kii} & $84.4$ & & $91.72$ & & $42.0(3)$ & $101.0(1.1)$ & \textit{171K} \\
        & \textbf{SParTA} & $\textit{85.06}$ & $85.04(1)$ & $92.02$ & $92.01(3)$ & $45.7(2)$ & $121.8(5)$ & 300k \\
        \cline{2-9}
        & ParticleNet-f.t.$_\text{exp}$ \cite{Qu:2022mxj} & $83.9$ & & $91.15$ & & $40.1(2)$ & $100.3(1.0)$ & \textit{370K} \\
        \textbf{(\textit{Fine-tuned})} & ParT-f.t.$_\text{full}$ \cite{Qu:2022mxj} & $85.2$ & & $92.30$ & & $50.6(2)$ & $138.7(1.3)$ & 2.14M \\
        & MIParT-L-f.t.$_\text{full}$ \cite{Wu:2024thh} & $\textit{85.3}$ & & $\textit{92.37}$ & & $\textit{51.9(5)}$ & $\textit{141.4(1.5)}$ & 2.38M \\
        & OmniLearn \cite{Mikuni:2025tar} & $84.4$ & & $91.59$ & & $43.7(3)$ & $107.7(1.5)$ & 1.6M \\
        \hline\hline
    \end{tabular}}
    \label{tab:qgTagRes1}
\end{table*}

\begin{table*}[!tb]
    \centering
    \small
    \renewcommand{\arraystretch}{1.25}
    \caption{\small Performance of jet taggers on the top-tagging test set. Conventions as in Table~\ref{tab:qgTagRes1}.}
    \setlength{\tabcolsep}{4pt}
    \resizebox{\textwidth}{!}{%
    \begin{tabular}{llccccccc}
    \hline\hline
        & & \multicolumn{2}{c}{Accuracy (\%)}  & \multicolumn{2}{c}{AUC (\%)} & \multicolumn{2}{c}{$1/\epsilon_B$} & \# of Params\\ 
        & & Best & Dist. & Best & Dist. & $\epsilon_S = 0.5$   & $\epsilon_S = 0.3$ & \\
        \hline
        & ParticleNet \cite{Qu:2019gqs} & $94.0$ & & $98.58$ & & $397(7)$ & $1615(93)$ & 370K \\
        & PFN \cite{Komiske:2018cqr} & -- & & $98.19$ & & $247(3)$ & $888(17)$ & \textbf{86.1K} \\
        & rPCN \cite{Shimmin:2021pkm} & -- & & $98.40$ & & $364(9)$ & $1642(93)$ & -- \\
        \hline
        \textit{Lorentz Invariance-} & PELICAN \cite{Bogatskiy:2023nnw} & $\textit{94.26}$ & & $\textit{98.70}$ & & -- & $\textit{2250(75)}$ & \textit{208K} \\
        \textit{based} & LorentzNet \cite{Gong:2022lye} & $94.2$ & & $98.68$ & & $498(18)$ & $2195(173)$ & 224K \\
        & L-GATr \cite{Brehmer:2024yqw} & $94.2$ & & $\textit{98.70}$ & & \textit{540(20)} & $2240(70)$ & 1.1M \\
        \hline\hline
        \textit{Attention-based} & PCT \cite{Mikuni:2021pou} & $94.0$ & & $98.55$ & & $392(7)$ & $1533(101)$ & 193.3K \\
        & ParT \cite{Qu:2022mxj} & $94.0$ & & $98.58$ & & $413(6)$ & $1602(81)$ & 2.14M \\
        \textbf{(\textit{Trained}} & Mixer \cite{Hammad:2024cae} & $94.0$ & & $98.59$ & & $416(5)$ & -- & \textit{86.03K} \\
        \textbf{\textit{from}} & MIParT \cite{Wu:2024thh} & $\textit{94.2}$ & & $98.68$ & & $505(8)$ & $2010(97)$ & 720.9K \\
        \textbf{\textit{scratch})} & IAFormer \cite{Esmail:2025kii} & $\textit{94.2}$ & & $\textit{98.70}$ & & $\textit{510(6)}$ & $\textit{2012(30)}$ & 211K \\
        & \textbf{SParTA}  & $94.1$ & $94.05(2)$ & $98.60$ & $98.60(1)$ & $443(8)$ & $1749(100)$ & 511K \\
        \cline{2-9}
        & ParticleNet-f.t. \cite{Qu:2022mxj} & $94.2$ & & $98.66$ & & $487(9)$ & $1771(80)$ & \textit{370K} \\
        \textbf{(\textit{Fine-tuned})} & ParT-f.t. \cite{Qu:2022mxj} & $94.4$ & & $98.77$ & & \textbf{\textit{691(15)}} & $2766(130)$ & 2.14M \\
        & MIParT-L-f.t. \cite{Wu:2024thh} & $94.4$ & & $98.78$ & & $640(10)$ & $2789(133)$ & 2.38M \\
        & L-GATr-f.t. \cite{Brehmer:2024yqw} & \textbf{\textit{94.46}} & & \textbf{\textit{98.793}} & & $651(11)$ & $\textit{2894(84)}$ & 1.1M \\
        & OmniLearn \cite{Mikuni:2025tar} & $94.2$ & & $98.72$ & & $568(9)$ & $2647(192)$ & 1.6M \\
        \hline\hline
    \end{tabular}}

    \label{tab:topTagRes1}
\end{table*}

\begin{table*}[t]
\centering
\caption{%
Parameters and FLOPs per jet (single forward pass) of transformer and
attention-based jet taggers. ``Architecture'' cites the model paper,
``Numbers'' the source of the values. FLOP counts depend on the number of
constituents and the counting method, so compare values only within the same
setting: \emph{JetClass, original}~\cite{Qu:2022mxj}; \emph{Top/QG}, MIParT
complexity table~\cite{Wu:2024thh}; \emph{$N=50$}, measured with PyTorch
\texttt{FlopCounterMode} for 50 constituents~\cite{Favaro:2025pgz, Petitjean:2025zjf, Favaro:2026amj};
\emph{JetClass 2M}, 2M-jet subset~\cite{Zheng:2026juf}; \emph{100 tokens,
analytic}, estimated as $2N+2\,n_{\rm attn}d_{\rm model}n_{\rm ctx}$
rather than measured~\cite{Esmail:2025kii}; \emph{100 particles}, up to 100
zero-padded constituents~\cite{Mikuni:2021pou}. $^\dagger$Optimized
implementation~\cite{Favaro:2026amj} of the original L-GATr; parameter count
unchanged. (b)~Joint scaling of Transformer, ParT, L-GATr(-slim) and
LLoCa-Transformer~\cite{Favaro:2026amj}; all agree within $\pm5\%$.
$^*$Lorentz-equivariant.}
\label{tab:flops_transformer_taggers}
\vspace{0.5em}
\textbf{(a)} Individual models\\[0.3em]
\small
\setlength{\tabcolsep}{4pt}
\resizebox{\textwidth}{!}{%
\begin{tabular}{l l r r l c}
\toprule
Model / variant & Architecture & Params & FLOPs & Setting & Numbers \\
\midrule
ParT                                   & \cite{Qu:2022mxj}       & 2.14M   & 340M   & JetClass, original      & \cite{Qu:2022mxj} \\
ParT (plain MHA)                       & \cite{Qu:2022mxj}       & 2.13M   & 260M   & JetClass, original      & \cite{Qu:2022mxj} \\
ParT                                   & \cite{Qu:2022mxj}       & 2.141M  & 211M   & $N=50$                  & \cite{Favaro:2025pgz} \\
LLoCa-ParT$^*$                         & \cite{Favaro:2025pgz}   & 2.160M  & 315M   & $N=50$                  & \cite{Favaro:2025pgz} \\
\midrule
MIParT                                 & \cite{Wu:2024thh}       & 720.9k  & 180M   & Top/QG                  & \cite{Wu:2024thh} \\
MIParT-L                               & \cite{Wu:2024thh}       & 2.38M   & 368M   & Top/QG                  & \cite{Wu:2024thh} \\
MIParT-L                               & \cite{Wu:2024thh}       & 2.380M  & 225M   & $N=50$                  & \cite{Favaro:2025pgz} \\
\midrule
Transformer (ParT-sized baseline)      & \cite{Favaro:2025pgz}   & 1.979M  & 210M   & $N=50$                  & \cite{Favaro:2025pgz} \\
LLoCa-Transformer$^*$                  & \cite{Favaro:2025pgz}   & 1.998M  & 301M   & $N=50$                  & \cite{Favaro:2025pgz} \\
LLoCa-Transformer$^*$ (narrow)         & \cite{Favaro:2025pgz}   & 1.980M  & 219M   & $N=50$                  & \cite{Petitjean:2025zjf} \\
\midrule
L-GATr$^*$ (original)                  & \cite{Spinner:2024hjm, Brehmer:2024yqw} & 1.079M & 2060M & $N=50$ & \cite{Favaro:2025pgz} \\
L-GATr$^*$ (optimized, sparse)         & \cite{Spinner:2024hjm, Brehmer:2024yqw} & 1.079M$^\dagger$ & 352M & $N=50$ & \cite{Favaro:2026amj} \\
L-GATr$^*$ (optimized, dense)          & \cite{Spinner:2024hjm, Brehmer:2024yqw} & 1.079M$^\dagger$ & 1999M & $N=50$ & \cite{Favaro:2026amj} \\
L-GATr-slim$^*$                        & \cite{Petitjean:2025zjf} & 2.031M & 329M   & $N=50$                  & \cite{Petitjean:2025zjf} \\
\midrule
JetFormer                              & \cite{Zheng:2026juf}    & 1.66M   & 213M   & JetClass 2M             & \cite{Zheng:2026juf} \\
\midrule
IAFormer (top tagging)                 & \cite{Esmail:2025kii}   & 211k    & 38M    & 100 tokens, analytic    & \cite{Esmail:2025kii} \\
Plain Transformer (IAFormer baseline)  & \cite{Esmail:2025kii}   & 2.1M    & 300M   & 100 tokens, analytic    & \cite{Esmail:2025kii} \\
\midrule
PCT                                    & \cite{Mikuni:2021pou}   & 193.3k  & 266M   & 100 particles           & \cite{Mikuni:2021pou} \\
SPCT                                   & \cite{Mikuni:2021pou}   & 7.0k    & 2.4M   & 100 particles           & \cite{Mikuni:2021pou} \\
\bottomrule
\end{tabular}}
 
\vspace{1em}
\textbf{(b)} Joint scaling family of Ref.~\cite{Favaro:2026amj} ($N=50$)\\[0.3em]
\resizebox{\textwidth}{!}{%
\begin{tabular}{l c c c c c}
\toprule
Size               & XS ($s=-2$) & S ($s=-1$) & M ($s=0$) & L ($s=1$) & XL ($s=2$) \\
\midrule
Channels / blocks / heads & 32 / 4 / 4 & 64 / 6 / 4 & 128 / 8 / 8 & 256 / 12 / 8 & 512 / 17 / 16 \\
Params ($\pm5\%$)  & 42k  & 250k & 1.4M & 7.9M & 42M \\
FLOPs ($\pm5\%$)   & 6M   & 30M  & 200M & 800M & 4G  \\
\bottomrule
\end{tabular}}
\end{table*}

\section{Conclusion}
\label{sec:concl}

We have introduced SParTA, a transformer for jet tagging that restricts its most expensive operation to the part of the jet where it matters most. The constituents are ordered by transverse momentum, and only the $N$ hardest of them, the hard core, enter a differential interaction attention whose weights are computed from pairwise physics features. The soft constituents are coupled to the hard core through cross-attention, whose cost grows only linearly with the multiplicity. The pairwise features are built inside the network from the constituent four-momenta and are conditioned on a vector of 53 jet-level observables, and a $\pt$-rank positional encoding gives the hard core a well-defined meaning.

A scan of 30 configurations on quark--gluon tagging showed which of these ingredients set the performance. Raising the depth from two to four blocks and enlarging the hard core from 60 to 80 constituents improved the AUC in every matched pair of configurations, while a sixth block added nothing and wider embeddings did not help. As a result, every Pareto-efficient configuration but one uses the narrowest embedding, and the hyperparameters that set the cost of the model are largely not those that set its discrimination. The high similarity between the representations of successive blocks supports the same picture of a network that saturates at moderate depth.

With 300k parameters, SParTA reaches an AUC of 92.02\% on quark--gluon tagging. This matches ParT, which has about seven times as many parameters, and improves on IAFormer by 0.30 points in AUC and by 21\% in background rejection at 30\% signal efficiency. On top tagging, SParTA outperforms ParT and ParticleNet but remains behind IAFormer, MIParT and the Lorentz-equivariant networks, with hyperparameters that were chosen on quark--gluon tagging and not re-optimised.

Several extensions follow directly from these results. The scan should be repeated with several seeds and with selection on the validation set, and a dedicated scan on top tagging should establish whether a deeper network or a different hard core closes the remaining gap. Training on the JetClass dataset would test how the architecture scales with data. Options that SParTA already supports but that were not used here, such as pointwise convolutions over the interaction tensor, class-attention aggregation and local neighbourhood features in the interaction embedding, can then be evaluated. Finally, a common protocol for counting FLOPs would allow the computational cost of SParTA to be compared directly with that of the networks in Table~\ref{tab:flops_transformer_taggers}.

\clearpage
\acknowledgments
RB, SM, and SKP would like to acknowledge support from ANRF, India (grant order no. CRG/2022/003208). SM would also like to acknowledge ANRF for grant order no. CRG/2023/008570. The authors also acknowledge help from Dr. Basabendu Barman for providing us with computational resources.

\section*{Code and data availability}
The trained network in multiple formats that are compatible with both Mathematica and PyTorch frameworks are publicly available at \href{https://github.com/BEC-HEP/SParTA}{\texttt{https://github.com/BEC-HEP/SParTA}}. The codes for the network and analysis will be made available on request.

\clearpage
\onecolumn
\appendix
\section{Architectural features of jet-tagging networks}
\label{app:features}
Table~\ref{tab:jet_tagging_feature_comparison} lists, for SParTA and seven other jet-tagging networks, which of the architectural features defined in Section~\ref{sec:architecture_feat} each one uses, grouped into symmetry and representation, pairwise interactions, attention and aggregation, and inputs, encoding and output. Architectural feature comparison is across eight jet-tagging networks: the subjet cross-attention Mixer-Network, the Attention-Based Cloud Network (ABCNet), the Edge Convolution Transformer (ECT), the Lorentz-Equivariant Geometric Algebra Transformer (L-GATr), the Particle Transformer (ParT), the More-Interaction Particle Transformer (MIParT), the Interaction-Aware Transformer (IAFormer), and SParTA (this work). Columns are grouped by backbone family --- MLP-mixer, graph-convolutional, and transformer --- and, within the transformers, by how pairwise information enters the attention.

\begin{table}[!h]
\centering
\footnotesize
\renewcommand{\arraystretch}{1.1}
\caption{\footnotesize 
AB: Additive Bias; AS: Attention Score; GC: Graph Convolution\\
\checkmark\ indicates the feature is present as described in the corresponding publication (and, where available, its reference implementation); -- indicates it is absent or not applicable. \\
$\dagger$ Feature not implemented in the current architecture, but is an option. \\
$*$ A convolution operation takes place when the interaction tensor is used in the Particle Attention Blocks.\\
$\sim$ denotes architectures that learn pairwise interaction features from \textit{embedded features} but do not combine them with the physical interaction features in a single tensor.\\
Marks refer to the architecture as published; several reference implementations expose additional options that are disabled in the published configuration. \\
The ``\textit{Activation in FFN}'' row reports the nonlinearity of the position-wise feed-forward sub-layer only; embedding-MLP and attention-coefficient activations may differ.}
\setlength{\tabcolsep}{2pt}
\begin{tabular}{>{\raggedright\arraybackslash}p{0.36\textwidth}*{8}{c}}
\hline
\textbf{Feature} &
\rotatebox{90}{\textbf{Mixer-Net}} &
\rotatebox{90}{\textbf{ABCNet}} &
\rotatebox{90}{\textbf{ECT}} &
\rotatebox{90}{\textbf{L-GATr}} &
\rotatebox{90}{\textbf{ParT}} &
\rotatebox{90}{\textbf{MIParT}} &
\rotatebox{90}{\textbf{IAFormer}} &
\rotatebox{90}{\textbf{SParTA}} \\
\hline
\multicolumn{9}{l}{\textit{Symmetry and representation}} \\
Exact Lorentz equivariance & -- & -- & -- & \checkmark & -- & -- & -- & -- \\
Longitudinal-boost-invariant pairwise interaction inputs & -- & -- & -- & -- & \checkmark & \checkmark & \checkmark & \checkmark \\
Geometric algebra (multivector) representation & -- & -- & -- & \checkmark & -- & -- & -- & -- \\
Tunable symmetry breaking (reference vectors) & -- & -- & -- & \checkmark & -- & -- & -- & -- \\
Permutation invariance & \checkmark & \checkmark & \checkmark & \checkmark & \checkmark & \checkmark & \checkmark & -- \\
\hline
\multicolumn{9}{l}{\textit{Pairwise interactions}} \\
Pairwise Interaction features used & -- & \checkmark & \checkmark & -- & \checkmark & \checkmark & \checkmark & \checkmark \\
Pairwise Interaction features used as & -- & AS & GC & -- & AB & AB & AS & AS \\
Pointwise convolutions over dense interaction tensor & -- & -- & -- & -- & \checkmark & \checkmark & \checkmark & \checkmark$^\dagger$ \\
Interaction embedding re-learned per layer & -- & -- & -- & -- & -- & \checkmark$^*$ & \checkmark & \checkmark \\
Interaction tensor built in-network from 4-momenta & -- & -- & -- & -- & \checkmark & \checkmark & -- & \checkmark \\
Learnable interaction channels beside physical features & -- & $\sim$ & $\sim$ & $\sim$ & -- & -- & -- & \checkmark \\
Global features injected into the interaction tensor & -- & -- & -- & -- & -- & -- & -- & \checkmark \\
\hline
\multicolumn{9}{l}{\textit{Attention and aggregation}} \\
Dot-product self-attention backbone & -- & -- & \checkmark & \checkmark & \checkmark & \checkmark & -- & \checkmark \\
Differential (two-softmax) sparse attention & -- & -- & -- & -- & -- & -- & \checkmark & \checkmark \\
Local neighbourhood graph convolution (EdgeConv / kNN) & -- & \checkmark & \checkmark & -- & -- & -- & -- & -- \\
Cross-attention to an auxiliary token set & \checkmark & -- & -- & -- & -- & -- & -- & \checkmark \\
Class-token / class-attention aggregation & -- & -- & \checkmark & -- & \checkmark & \checkmark & -- & \checkmark \\
Pooling-based aggregation & \checkmark & \checkmark & -- & \checkmark & -- & -- & \checkmark & \checkmark \\
Hard/soft particle sparsity split & -- & -- & -- & -- & -- & -- & -- & \checkmark \\
\hline
\multicolumn{9}{l}{\textit{Inputs, encoding and output}} \\
Explicit jet-level (global) features & -- & \checkmark & \checkmark & -- & -- & -- & -- & \checkmark \\
Explicit rank / positional embedding & -- & -- & -- & -- & -- & -- & -- & \checkmark \\
Per-particle (node-level) output head & -- & \checkmark & -- & -- & -- & -- & -- & -- \\
Activation in FFN & GELU & ReLU & ReLU & gated GELU & GELU & GELU & SiLU & SwiGLU \\
\hline
\end{tabular}

\label{tab:jet_tagging_feature_comparison}
\end{table}

\clearpage

\bibliographystyle{JHEP}
\bibliography{qgl}

\end{document}